\documentclass[	a4paper,
 reprint,
 superscriptaddress,
 amsmath,amssymb,
 aps,
 prd,
 nofootinbib
]{revtex4-2}

\clearpage{}\usepackage[uline]{hhtensor}
\usepackage{mathrsfs}
\usepackage{braket}
\usepackage{hyperref}
\usepackage{mathtools}
\usepackage{siunitx}
\usepackage{xcolor}
\usepackage{textgreek}
\usepackage{framed}
\usepackage{empheq}

\newcommand\ff[1]{\mathrm{f}_{#1}(\pp)}

\newcommand\ffe[1]{f_{#1}(\pp)}
\newcommand\ffestar[1]{f^*_{#1}(\pp)}
\newcommand\FFe[1]{f_{jm#1}(k)}
\newcommand\FFestar[1]{f^*_{jm#1}(k)}

\newcommand*{\hbaraux}[2]{\sbox0{\mathsurround=0pt$#1\mkern-1mu\mathchar'26$}\mkern-1mu\lower.07\ht0\box0\mkern-8mu}
\renewcommand*{\hbar}{{\mathpalette\hbaraux\relax\mathrm{h}}}

\newcommand\Eq[1]{\text{Eq.~(\ref{#1})}}

\newcommand\rr{\mathbf{r}}

\newcommand\pp{\mathbf{k}}

\newcommand\phat{\mathbf{\hat{\pp}}}

\newcommand\Ert{\mathbf{E}(\rr,t)}

\newcommand\intdpinv{\int_{\mathbb{R}^3} \frac{\text{d}^3 \pp}{k}\text{ }}
\newcommand\intdpinvmz{\int_{\mathbb{R}^3-\zerovec} \frac{\text{d}^3 \pp}{k}\text{ }}

\newcommand\zerovec{\mathbf{0}}

\newcommand\ii{\mathrm{i}}

\newcommand\epsz{\text{\textepsilon}_{\text{0}}}
\newcommand\cz{\mathrm{c_0}}

\newcommand\sss{\mathbf{s}}

\newcommand\rhat{\mathbf{\hat{r}}}

\DeclareMathOperator{\atantwo}{atan2}

\newcommand\Id{\mathrm{I}}
\newcommand\op[1]{\mathrm{#1}}

\newcommand\barlambda{\bar{\lambda}}

\newcommand\qq{\mathbf{q}}

\newcommand\intdkmeasure{\int_{>0}^\infty {\mathrm{d}k}k}
\newcommand\intdkZmeasure{\int_{0}^\infty {\mathrm{d}k}k}

\newcommand\intdqmeasure{\int_{>0}^\infty {\mathrm{d}q}q}

\newcommand\jbar{\bar{j}}

\newcommand\mbar{\bar{m}}

\newcommand{\sandwich}[3]{\langle#1|#2|#3\rangle}
\newcommand\inout{\text{in/out}}
\providecommand\M{\mathbb{M}}

\providecommand\anj{\sqrt{\frac{4(n-j-1)!}{(n+j)!}}}
\providecommand\anjpw{\sqrt{\frac{(2j+1)(n-j-1)!}{\pi(n+j)!}}}

\providecommand\nk{n_{\text{Kastrup}}}

\clearpage{}
\usepackage[thinc]{esdiff}
\usepackage{booktabs}
\begin{document}
\title{Countable basis for free electromagnetic fields}
\author{Ivan Fernandez-Corbaton}
\email{ivan.fernandez-corbaton@kit.edu}
\affiliation{Institute of Nanotechnology, Karlsruhe Institute of Technology, Kaiserstr. 12, 76131 Karlsruhe, Germany}
\begin{abstract}
	Polychromatic electromagnetic fields are typically expanded as integrals over monochromatic fields, such as plane waves, multipolar fields, or Bessel beams. However, monochromatic fields do not belong to the Hilbert space of free Maxwell fields, since their norms diverge. Moreover, the continuous frequency integrals involved in such expansions complicate the treatment of light--matter interactions via the scattering operator. Here, we identify and study a polychromatic basis for free Maxwell fields whose basis vectors belong to the Hilbert space. These vectors are defined as simultaneous eigenstates of four commuting operators with integer eigenvalues. As a consequence, the basis set is countable, and the Hilbert space is separable and isomorphic to $\ell^2$, the Hilbert space of square-summable sequences. Each basis vector represents a polychromatic single-photon wave with quantized energy and a wavelet--like temporal dependence. Three versions of this basis are defined: Regular, incoming, and outgoing. The fields of the regular basis are smooth in both space and time. The incoming and outgoing fields are likewise smooth, except at the spatial origin. These results support and motivate the use of countable bases for both the theoretical description and the practical computation of light--matter interactions.
\end{abstract}
\maketitle
\section{Introduction and summary}
The Hilbert space of free Maxwell fields has been studied since the 1960's \cite{Lomont1964,Kastrup1966,Moses1965,Moses1965b,Moses1967,Mack1969,Moses1973b,Birula1975,Birula1996,Moses2004,Kastrup2008,Todorov2019}, often in connection to symmetry groups of Maxwell equations, such as the Poincar\'e or conformal groups. Such a Hilbert space, which we denote by $\M$, supports the modeling of light-matter interactions by means of the scattering operator $S$. The scattering operator of a material system maps any incident illumination, defined as an incoming free field prior to interaction, to the resultant outgoing free field radiated by the system once radiation from the matter has ceased. The formulation of electromagnetism in the Hilbert space has theoretical and practical benefits. For example, fundamental properties of any given field such as number of photons, energy, and momentum, can be expressed by means of scalar products \cite{Zeldovich1965,Birula1996}. The review in \cite{FerCor2024b} contains an overview of the main theoretical ideas, and some of the practical benefits in current developments.

The relationship $S=\Id+T$ connects the scattering operator to the T-matrix, which is a popular and powerful computational framework \cite{Waterman1965,Gouesbet2019,Mishchenko2020}. While the T-matrix was typically restricted to mappings between monochromatic fields, which do not belong in $\M$ because their norm diverges, a recent polychromatic reformulation of the framework \cite{Vavilin2023} has connected $\M$ with the T-matrix. The polychromatic framework allows one to systematically treat the interaction of general pulses of light with objects, including those moving at relativistic speeds \cite{Vavilin2024,Whittam2024}. Other recent work \cite{Ambrosio2024} has extended the generalized Lorenz--Mie theory {\em stricto sensu}, which applies to spherical objects, from monochromatic to polychromatic fields.

The polychromatic formulation in $\M$ has room for improvement. The fields in $\M$ are expanded by integrals over monochromatic fields which are not in $\M$. Moreover, monochromatic fields are not physically meaningful. For example, their total energy is not well-defined because their energy density integrated over the whole space diverges to infinity. The situation is rather awkward from the theoretical point of view: A proper basis with elements in $\M$ is missing. Also, the expressions of the $S$ and $T$ operators feature a double frequency integral, which reduces to a single frequency integral in some cases. Such integrals require the discretization of continuous variables, which is sometimes cumbersome \cite{Vavilin2024}, and generally increase the complexity of practical calculations with respect to the much simpler vector-matrix multiplications in the traditional monochromatic setting.

Addressing these issues requires one to consider fundamental aspects of $\M$, namely its dimension and the existence of a countable basis. The dimension of a Hilbert space is the number of vectors in an orthonormal basis. The dimension of $\M$ must be infinite so that it matches with the infinite number of degrees of freedom of the field. One may then ask whether such an infinite is countable of uncountable. An infinite countable set is a set whose elements can be counted by the natural numbers: There exists a one-to-one correspondence between the set and the natural numbers. The existence of a countable basis in a Hilbert space is important. Hilbert spaces with a countable basis are called separable. Separability has been assumed in the axiomatic foundations of quantum mechanics \cite[p.~46]{VonNeumann1955}, and quantum field theory \cite[Sec.~3.1]{Streater1989}. Removing the separability assumption brings about problems such as questionable vacuum states \cite{Earman2020}. The existence of a countable basis makes any Hilbert space isomorphic to $\ell^2$, the Hilbert space of square-summable sequences, which is the prototypical Hilbert space whose infinite degrees of freedom are manageable, precisely thanks to the existence of a countable basis.

None of the sets of fields that are typically used for expanding Maxwell fields is countable, because they are defined as simultaneous eigenstates of at least one operator with a continuous non-countable spectrum. For plane waves, the linear momentum operators $\op{P}_{x}$, $\op{P}_{y}$, and $\op{P}_{z}$ have continuous spectra, as do the energy operator $\op{H}$ and $\op{P}_z$ for Bessel beams, and $\op{H}$ for multipolar fields. Moreover, plane waves, multipolar fields and Bessel beams {\em are not} members of $\M$, because, as is the case for all monochromatic fields, their norm diverges. Actually, the fact that monochromatic fields are not members of $\M$ is reassuring in the following sense: Since they do not form a proper orthonormal basis in $\M$, one cannot conclude that the dimension of $\M$ is infinitely uncountable.

We will see that the electromagnetic Hilbert space is separable, that is, it has a countable basis. This can be deduced by extrapolating the basis provided in \cite{Kastrup1966,Kastrup1970} by Kastrup and Mayer for the scalar wave equation, or by particularizing the more general basis provided by Mack and Todorov \cite{Mack1969} to electromagnetic fields. The two bases are defined using a different set of operators, and the expressions in \cite{Mack1969,Kastrup1966,Kastrup1970} are dimensionally inconsistent and not appropriate for practical use.

In here, we unify two apparently different countable basis for $\M$ and identify and resolve dimensional inconsistencies in their expressions. The new expressions of the expansion functions of the basis vectors in plane waves and multipolar fields, which can be found in \Eq{eq:pwcoeffsnew} and \Eq{eq:mpcoeffsnew}, respectively, are ready to use in practical applications. The basis vectors are defined by four integers $\ket{njm\lambda}$, where $j$ and $m$ define the total angular momentum square and the angular momentum along the $z$ axis, respectively, $\lambda$ defines the helicity or polarization handedness, and $n$ is the eigenvalue of a self-adjoint operator, sometimes referred to as a ``conformal Hamiltonian''. Each $\ket{njm\lambda}$ represents a single polychromatic photon with energy equal to $n\hbar\cz k_0$, where $k_0^{-1}$ is a length scale that can be chosen to fit the problem at hand. We show that the fields of the $\ket{njm\lambda}$ are smooth in space and time, that is, one can differentiate them with respect to $x,y,z,$ and $t$ an arbitrary number of times. The analysis of their time dependence reveals a remarkable a double wavelet-character. One of the wavelets is essentially of incoming character and the other of outgoing character. Taking only one of them allows one to define countable bases for incoming or outgoing fields, which are also smooth in space and time except at the origin of coordinates $\rr=[0,0,0]$. The countable basis was called {\em the canonical basis} in \cite{Mack1969}, because it is defined in terms of the eigenvectors of the maximal set of commuting operators with a discrete spectrum. The basis for $\M$ obtained in this article can also be called {\em canonical} in the same sense.

The rest of the article is organized as follows. Section~\ref{sec:mconf} reviews the electromagnetic Hilbert space $\M$ and its connection with the conformal group. In Sec.~\ref{sec:two}, two existing and apparently different countable basis for $\M$ are shown to be the equivalent because one such basis involves a particular self-adjoint operator $N$, and the other involves the unitary transformation generated by $N$. Dimensional discrepancies in the existing expressions for the basis vectors are identified and resolved in Sec.~\ref{sec:newexp}. Section~\ref{sec:prop} contains a study of some of the properties of the electromagnetic fields of the basis vectors, namely: That they have zero static component, that they are smooth in space and time, and that they have a double wavelet time-dependence, one of incoming character and one outgoing character. Finally, Sec.~\ref{sec:con} contains the conclusions and outlook.

\section{The electromagnetic Hilbert space and the conformal group\label{sec:mconf}}
Let us consider an electromagnetic field after its radiation by a material system is finished and before it starts interacting with another material system. We call such a field a free field. The set of free electromagnetic fields, together with the electromagnetic scalar product \cite{Gross1964}, form a Hilbert space. The scalar product between two members of $\mathbb{M}$ can be computed as \cite{Vavilin2023}:
\begin{equation}
	\label{eq:lmsp}
	\begin{split}
		\langle f|g\rangle &= \sum_{\lambda=\pm1} \int_{\mathbb{R}^3-\zerovec} \frac{\text{d}^3 \pp}{k} \, \ffestar{\lambda}g_\lambda (\pp)\\
		&=\sum_{\lambda=\pm 1} \int_{>0}^{\infty} \text{d}k\, k \sum_{j=1}^{\infty} \sum_{m=-j}^j \, \FFestar{\lambda} {g}_{jm\lambda}(k),
	\end{split}
\end{equation}
where $\pp$ is the wavevector, $k=|\pp|$ the wavenumber, $\lambda$ the polarization handedness, and $\ff{\lambda}$ and $\mathrm{g}_\lambda (\pp)$ are the plane wave expansion functions of $\ket{f}$ and $\ket{g}$, respectively. The $\FFe{\lambda}$ and ${g}_{jm\lambda}(k)$ are the corresponding expansion functions in spherical waves, also known as multipolar fields. The integer $j=1,2,...$ is the multipolar degree, with $j=1$ corresponding to dipoles, $j=2$ to quadrupoles, and so on, and the integer $m=-j,-j+1,\ldots, j$ is the component of the angular momentum in the $z$ direction. Removing the point at $|\pp|=k=0$ \cite{Gross1964} avoids undesired contributions from static fields. 

The form of the expressions in \Eq{eq:lmsp} is reached using the conventions established in \cite{Vavilin2023} for the definitions of plane waves and multipolar fields, and for the corresponding expansions of the electromagnetic fields (see App.~\ref{app:m}). With these conventions, the expansion functions have units of meters:
\begin{equation}
	\label{eq:meters}
	\left[f_\lambda(\pp)\right]=\left[f_{jm\lambda}(k)\right]=\text{m},
\end{equation}
and, crucially, the plane waves, multipolar fields, corresponding expansion functions, and general states, transform as massless unitary irreducible representations of the Poincar\'e group with $|\lambda|=1$, as prescribed by Wigner \cite{Wigner1939}.

The scalar product provides direct connections from $\M$ to physically relevant quantities. The number of photons is simply the norm squared \cite{Zeldovich1965} $\braket{f|f}$. The amount of fundamental quantities contained in a given field is easily formulated as $\sandwich{f}{\Gamma}{f}$, where $\Gamma$ is the self-adjoint operator representing a given fundamental quantity, such as energy or momentum. For example, in the multipolar basis, the energy of the field is:
\begin{equation}
	\label{eq:Hen}
	\sandwich{f}{\op{H}}{f}=\sum_{\lambda=\pm 1} \sum_{j=1}^{\infty} \sum_{m=-j}^j \intdkmeasure \left(\hbar\cz k\right)|\FFe{\lambda}|^2.
\end{equation}

The self-adjoint operators $\Gamma$ representing the fundamental quantities generate unitary transformations in $\M$ through their exponentials: 
\begin{equation}
	\label{eq:exp}
	X(\gamma)=\exp\left(-\frac{\ii \gamma}{\hbar} \Gamma\right)=\sum_{s=0}^\infty\frac{\left(-\frac{\ii \gamma}{\hbar} \Gamma\right)^s}{s!}.
\end{equation}
For example, angular momentum operators $\op{J}_{\{x,y,z\}}$ generate rotations, and linear momentum operators $\op{P}_{\{x,y,z\}}$ generate translations.

At the beginning of the twentieth century Bateman and Cunningham showed \cite{Bateman1910,Cunningham1910} that Maxwell equations are invariant under a group larger than the Poincar\'e group, namely the 15 parameter conformal group in 3 dimensions of space, and 1 dimension of time: $C_{15}(3,1)$. The generators of $C_{15}(3,1)$ and their unitary transformations are, with $\mu\in\{0,1,2,3\}$ and $l\in\{1,2,3\}$: 
\begin{equation}
	\label{eq:generators}
	\begin{split}
		\op{P_\mu}&\text{ generate space-time translations,}\\
		\op{J_{l}}&\text{ generate spatial rotations,}\\
		\op{L_{l}}&\text{ generate Lorentz boosts,}\\
		\op{D}&\text{ generates dilatations: }x_\mu \rightarrow \alpha x_\mu\text{ with real }\alpha\ge 0,\text{ and}\\
		\op{K_\mu}&\text{ generate special conformal transformations:}\\
		x_\mu&\rightarrow\frac{x_\mu-b_{\mu} x^\mu x_\mu}{1-2 b^\mu x_\mu + b^\mu b_\mu x^\mu x_\mu}.
	\end{split}
\end{equation}

Together, the generators in the first three lines of \Eq{eq:generators} form the Lie algebra of the Poincar\'e group. The physical meaning of the transformations that they generate is well-known. The dilatations generated by $\op{D}$ are readily understood as homogeneous expansions or shrinkages of space-time. The interpretation of the special conformal transformations is less straight-forward. Kastrup argued that {\em they are not} changes to accelerating reference frames, and put forward their interpretation as space-time dependent changes of measurement units \cite{Kastrup1962}. Such interpretation is very intuitive when applied to dilatations, where the length of the measuring sticks is changed uniformly across space-time.

\section{Two countable basis for $\M$ are equivalent\label{sec:two}}
Mack and Todorov \cite{Mack1969} connected the conformal and Poincar\'e groups for massless fields, showing that the irreducible representations in the Poincar\'e group had corresponding irreducible representations in the conformal group. Moreover, they defined a countable basis for massless fields of any helicity, which they called the canonical basis \cite[Sec.~4]{Mack1969}. 

The elements of the countable basis in \cite[Sec.~4]{Mack1969}, $\ket{njm\lambda}$ are characterized by their eigenvalues with respect to four commuting operators. The four operators have a discrete spectrum labeled by integers \footnote{The factors of $\hbar$ in the last three lines of \Eq{eq:comops} do not appear in \cite{Mack1969} because they are implicitly set to 1. $N$ is defined in the first line for conciseness, and the $\hbar$ has been added there to match the convention in \Eq{eq:exp} for generating a unitary transformation from a self-adjoint operator.}:
\begin{equation}
	\label{eq:comops}
	\begin{split}
		\hbar\frac{\op{P_0}+\op{K_0}}{2}\ket{njm\lambda}&=N\ket{njm\lambda}= \hbar n\ket{njm\lambda},\ n=2,3,\ldots\\
		\op{J}^2\ket{njm\lambda}&=\hbar^2j(j+1)\ket{njm\lambda},1\le j\le n-1\\
		\op{J_z}\ket{njm\lambda}&=\hbar m\ket{njm\lambda},\ -j\le m\le j\\
		\op{\Lambda}\ket{njm\lambda}&=\hbar \lambda\ket{njm\lambda},\ \lambda\in\{-1,+1\}.
	\end{split}
\end{equation}
The operators $\op{J_z}$, and $\op{J}^2 =\left(\op{J_x}^2+\op{J_y}^2+\op{J_z}^2\right)$ characterize the rotational properties. The helicity operator:
\begin{equation}
	\Lambda=\frac{\op{\mathbf{J}}\cdot\op{\mathbf{P}}}{|\op{\mathbf{P}}|},
\end{equation}
generates the electromagnetic duality transformation \cite{Calkin1965,Zwanziger1968}. The eigenstates of helicity are fields of pure circular polarization handedness: Left-handed for $\lambda=1$, and right-handed for $\lambda=-1$. All the plane waves contained in the decomposition of a helicity eigenstate are of one and the same circular polarization handedness. Helicity commutes with all the generators in \Eq{eq:generators}, and is essentially one of the Casimirs of the conformal group. Since electromagnetic duality is a symmetry of free Maxwell fields, one can add it to $C_{15}(3,1)$ and enlarge the symmetry group to a 16-parameter Lie group\footnote{I am not aware of this natural enlargement being present in the literature. I first heard the idea from Konstantin Bliokh during a private conversation.}.

The operator $N$ has been shown to have a discrete positive spectrum bounded from below \cite{Luescher1975}. It is one of the operators referred to as ``conformal Hamiltonian'', whose role is somewhat analogous to the notion of the energy operator, which generates time translations, except that the ``time'' connected to $N$ is a coordinate in a compactification of the Minkowski space-time \cite{Todorov2019}. We will see below that $N$ is also the generator of the conformal inversion, which is an important conformal transformation because one can generate the whole Lie algebra of the group with the conformal inversion and the linear momentum operators \cite{Kastrup2008}.

Kastrup and Mayer obtained a countable basis for solutions of the scalar wave equation \cite{Kastrup1966,Kastrup1970}. The basis vectors are eigenstates of $\op{J}^2$, $\op{J}_z$. Helicity is not mentioned in \cite{Kastrup1966,Kastrup1970}, but the scalar case implies $\lambda=0$, which is actually implicit in the scalar spherical harmonics contained in the expression of their basis vectors (e.g. \cite[Eq.~(12)]{Kastrup1970}), through the relationship:
\begin{equation}
	Y_{jm}(\phat)=\sqrt{\frac{2j+1}{4\pi}}\exp(\ii m\phi)d_{m(\lambda=0)}^j(\theta).
\end{equation}

The fourth operator is, instead of $(\op{P}_0+\op{K}_0)/2$, the conformal inversion operator:
\begin{equation}
	\label{eq:R}
	\op{R}: x_\mu\rightarrow -\frac{x_\mu}{ x^\mu x_\mu}.
\end{equation}
While the involvement of two different operators suggests that the two countable basis are fundamentally different, the similarity of their expressions, which can be appreciated in Table~\ref{tab:expressions} of App.~\ref{app:pexp}, suggest otherwise. This apparent contradiction is resolved by realizing that $\op{R}$, which is sometimes called Weyl reflection, can be obtained as a ``rotation'' generated by $N$ \cite[below Eq.~(3.3)]{Budinich1993}:
\begin{equation}
	\label{eq:Rd}
	\op{R}=\left.\exp\left(-\frac{\ii}{\hbar} \theta N\right)\right|_{\theta=\pi}=\exp\left(-\ii\pi \left[\frac{\op{P}_0+\op{K}_0}{2}\right]\right).
\end{equation}
Then: 
\begin{equation}
\op{R}\ket{njm\lambda}=\exp(-\ii \pi n)\ket{njm\lambda}=(-1)^n\ket{njm\lambda},
\end{equation}
which is the eigenvalue of $R$ that appears in \cite{Kastrup1966,Kastrup1970}.

Therefore, the canonical basis defined by Mack and Todorov, when particularized to the scalar case, coincides with the basis defined by Kastrup and Mayer. In this respect, the minus sign in \Eq{eq:R} is actually crucial. If, instead, one uses another popular version of the conformal inversion $\op{\tilde{R}}: x_\mu\rightarrow x_\mu/(x^\mu x_\mu)$, then \Eq{eq:Rd} is not valid anymore for $\op{\tilde{R}}$, which is not connected to the identity, while $\op{R}$ is connected in the sense that $\lim_{\theta\rightarrow 0}\exp\left(-\ii\theta/\hbar N\right)=\Id$. Another consequence of the different sign is that, while $\op{R}$ does not affect helicity because its generator $N$ commutes with $\Lambda$, helicity changes sign under the action of $\op{\tilde{R}}$: $\lambda\rightarrow -\lambda$. The basis of common eigenstates cannot be defined with $\op{\tilde{R}}$. The helicity change due to $\op{\tilde{R}}$ can be deduced in several ways. One way is using results obtained in the context of Robinson congruences and null electromagnetic fields, where it is found that \cite[Sec.~5.4]{Dalhuisen2014},\cite[Eq.~(197)]{Arrayas2017}: $\mathbf{E}+\ii \cz \mathbf{B}\stackrel{\op{\tilde{R}}}{\rightarrow}\mathbf{\bar{E}}-\ii \cz \mathbf{\bar{B}}$. This change of sign implies the change of helicity since, in the complex representation of electromagnetic fields, $\mathbf{E}+\lambda\ii \cz \mathbf{B}$ are the eigenstates of $\Lambda$ with eigenvalue $\lambda$ \cite[Sec.~2.2.1]{FerCorTHESIS}. The same change of sign appears when using the action of $\op{\tilde{R}}$ on electromagnetic fields reported by Cunningham \cite[p.~89]{Cunningham1910}. The helicity flip can also be readily deduced by considering that $\op{\tilde{R}}$ is obtained from applying $\op{R}$, time reversal $t\rightarrow -t$, and parity $\rr\rightarrow -\rr$, and that helicity is unchanged by time reversal and flipped by parity. 

\section{New expressions for the countable basis\label{sec:newexp}}
Despite having proved their equivalence, the direct comparison of the countable bases given in \cite{Mack1969} and \cite{Kastrup1970} is not straightforward. Appendix~\ref{app:pexp} contains their explicit expressions and several clarifications regarding the different conventions that were used in those works. Importantly, there is a more fundamental problem: The expressions in \cite{Mack1969,Kastrup1966,Kastrup1970} are dimensionally inconsistent, and hence not appropriate for practical use. We will resolve this problem in this section.
\subsection{Dimensional discrepancies in the existing expressions}
Let us now consider the plane-wave expansion of $\ket{njm\lambda}$, which can be deduced from \cite[Eq.~(4.4)]{Mack1969} (see App.~\ref{app:pexp}):, and reads as:
\begin{equation}
	\label{eq:macks}
	\begin{split}
		&c_{njm\lambda}(\pp)=\langle \lambda \pp|njm\lambda\rangle=\anjpw\times\\
		&\exp(-k)(2k)^{j}L^{2j+1}_{n-j-1}(2k)\exp(\ii m\phi)d^j_{m\lambda}(\theta),
	\end{split}
\end{equation}
where the wavevector $\pp$, with $|\pp|=k$, is defined using angles $\theta\in[0,\pi)$ and $\phi\in[-\pi,\pi)$: 
\begin{equation}
	\pp=\left[k\sin(\theta)\cos(\phi),k\sin(\theta)\sin(\phi),k\cos(\theta)\right],
\end{equation}
$L^s_n(\cdot)$ denote the Laguerre polynomials defined by:
\begin{equation}
	L^s_l(\rho)=\sum_{r=0}^l\binom{l+s}{l-r}\frac{(-\rho)^r}{r!},
\end{equation}
and the $d^l_{qs}(\cdot)$ are the small Wigner d-matrices as defined in \cite[Chap.~7]{Tung1985}.

With the relation between $\ket{\pp\lambda}$ and $\ket{kjm\lambda}$ (in e.g. \cite[Eq.~(40)]{Vavilin2023}), one can readily reach the expression for the multipolar expansion functions of $\ket{njm\lambda}$:
\begin{equation}
	\label{eq:mpcoeffs}
	\begin{split}
		c_{nj}(k)&=\langle \lambda mj k\ket{njm\lambda}=\\
		&\anj\exp(-k)(2k)^{j}L^{2j+1}_{n-j-1}(2k),
	\end{split}
\end{equation}
finding that the expansion functions do not depend on $m$ or $\lambda$. Essentially equivalent expressions were found in \cite{Kastrup1966,Kastrup1970} for the scalar case (see e.g. \cite[Eq.~(12)]{Kastrup1966}).

It is worth mentioning here a known remarkable fact: The $\pp$ dependence of the plane wave expansion functions of $\ket{njm\lambda}$ in \Eq{eq:macks} is, for the scalar case, essentially identical to the $\rr$ dependence of the wavefunctions of the bound states of the hydrogen atom. This amazing connection is due to the role of the conformal group in generating the quantum mechanical states of the hydrogen atom \cite{Barut1967}. 

With the above results, \Eq{eq:xpans}, and \Eq{eq:ypans}, one can write the expansion of the $\ket{njm\lambda}$ in plane waves:
\begin{equation}
	\begin{split}
	&\ket{njm\lambda}=\intdpinvmz c_{njm\lambda}(\pp)\ket{\pp\lambda}\\
		&=\intdpinvmz \anjpw\times\\
		&\exp(-k)(2k)^{j}L^{2j+1}_{n-j-1}(2k)\exp(\ii m\phi)d^j_{m\lambda}(\theta)\ket{\pp\lambda},
	\end{split}
\end{equation}
and in multipolar fields:
\begin{equation}
	\label{eq:nmp}
	\begin{split}
		\ket{njm\lambda}&=\intdkmeasure c_{nj}(k)\ket{kjm\lambda}=\intdkmeasure\\
		&\anj\exp(-k)(2k)^{j}L^{2j+1}_{n-j-1}(2k)\ket{kjm\lambda}.
	\end{split}
\end{equation}
At this point, and referring back to \Eq{eq:meters} and the discussions around it, one can identify some dimensional discrepancies. The expansion functions $c_{njm\lambda}(\pp)$ and $c_{njm\lambda}(k)$ must have units of meters, but such dimensions do not seem to appear in \Eq{eq:macks}. Moreover, the $k$ in the arguments of the functions in \Eq{eq:macks} must be unitless, but a wavenumber has units of inverse meters. The expressions in \cite{Kastrup1966,Kastrup1970} have the same problems.

\begin{figure*}[t]
  \centering
  \begin{minipage}[t]{0.48\textwidth}
	  \vspace{0pt}
    \centering
    \includegraphics[width=\linewidth]{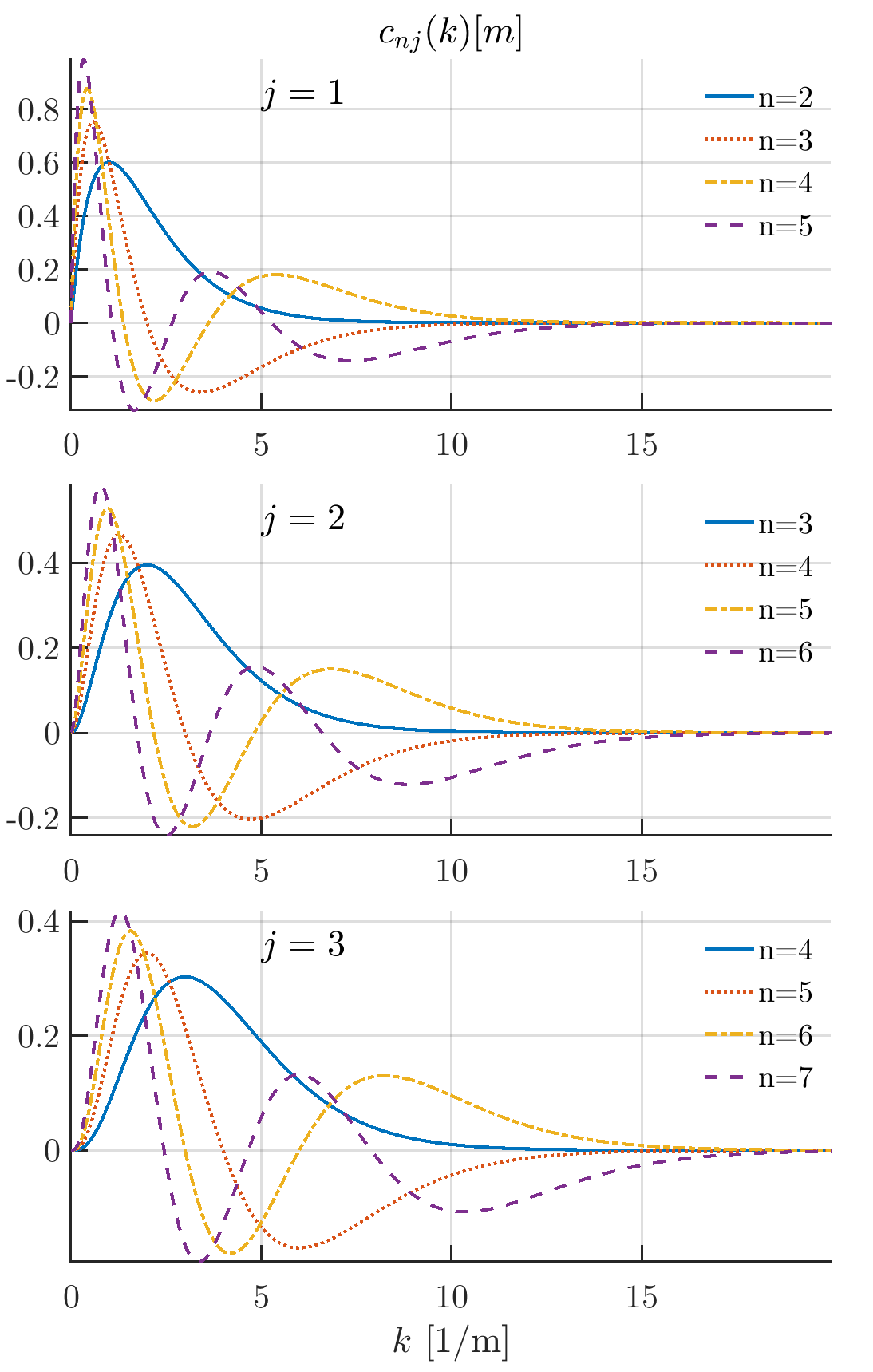}
  \end{minipage}\hfill
  \begin{minipage}[t]{0.48\textwidth}
	  \vspace{2pt}
    \centering
       
        		        \includegraphics[width=\linewidth]{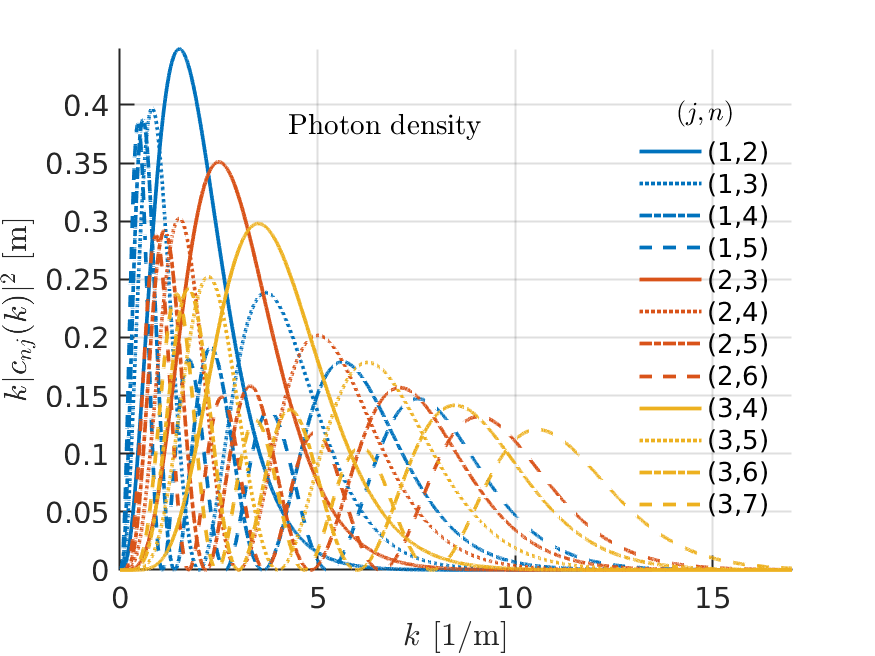}\\[0.5em]
    \includegraphics[width=\linewidth]{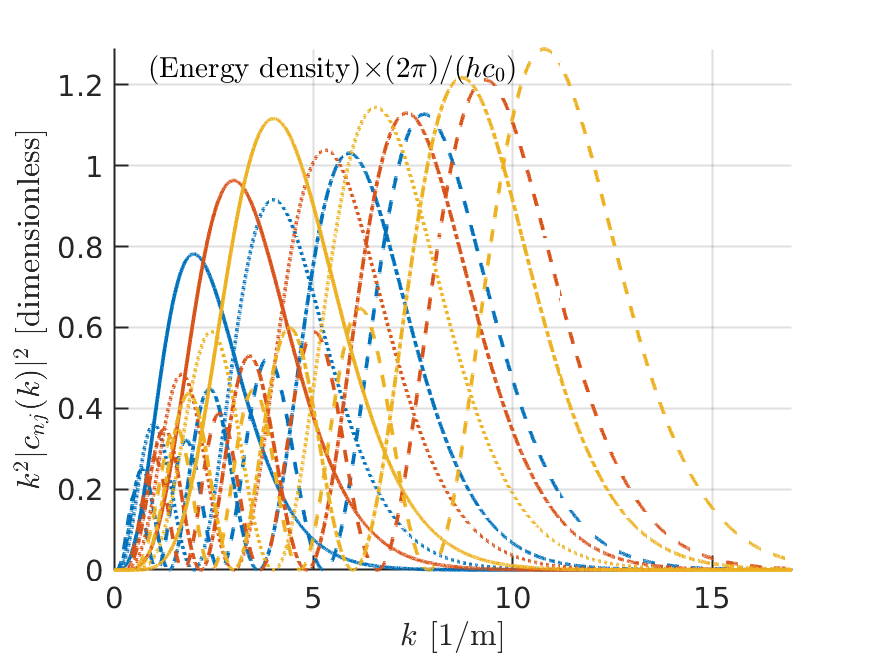}
  \end{minipage}
	\caption{\label{fig:enj}Left panels: Multipolar expansion functions $c_{nj}(k)$ of some members of the countable basis $\ket{njm\lambda}$ for the $j=1,2$ and $3$ multipolar orders, and $k_0=\SI{1}{\per\meter}$. Right panels (shared legend): Spectral densities ($\text{d}k$) of the same basis vectors. Top right: Photon density. Bottom right: Energy density.}
\end{figure*}

\subsection{Resolution of the discrepancies\label{sec:resolution}}
The discrepancies identified above can be solved using a quantity with units of length, which we choose to be the inverse of a wavenumber $k^{-1}_0$. We then write \Eq{eq:macks} as: 
\begin{equation}
	\label{eq:pwcoeffsnew}
	\boxed{
	\begin{split}
		&c_{njm\lambda}(\pp)=\langle \lambda \pp|njm\lambda\rangle=\anjpw\times\\
		&\frac{\exp(-k/k_0)}{k_0}(2k/k_0)^{j}L^{2j+1}_{n-j-1}(2k/k_0)\exp(\ii m\phi)d^j_{m\lambda}(\theta),
	\end{split}
}
\end{equation}
which changes \Eq{eq:mpcoeffs} into:
\begin{equation}
\label{eq:mpcoeffsnew}
	\boxed{
	\begin{split}
		&c_{nj}(k)=\langle \lambda mj k\ket{njm\lambda}=\\
		&\anj\frac{\exp(-k/k_0)}{k_0}(2k/k_0)^{j}L^{2j+1}_{n-j-1}(2k/k_0).
	\end{split}
}
\end{equation}
Additionally, we re-write the first line of \Eq{eq:comops} as:
\begin{equation}
	N=\hbar\left(\frac{P_0}{k_0}+k_0K_0\right),\, N\ket{njm\lambda}=\hbar n\ket{njm\lambda},
\end{equation}
which also affects \Eq{eq:Rd}:
\begin{equation}
	\label{eq:Rdd}
	\op{R}=\left.\exp\left(-\frac{\ii}{\hbar} \theta N\right)\right|_{\theta=\pi}=\exp\left(-\ii\pi \left[\frac{\op{P}_0/k_0+k_0\op{K}_0}{2}\right]\right).
\end{equation}
For the original equations to match the dimensional requirements, one needs to interpret them as being the new equations with the implicit choice of $k_0=\SI{1}{\meter^{-1}}$. 

The introduction of a quantity with units of length is not arbitrary. There is a radius implicitly assumed to be $r_0=\SI{1}{\meter}$ in the conformal inversion operator $R$. The inversion as written in \Eq{eq:R} maps all the points inside the unit hyperboloid to its outside, and vice versa. The operation $x_\mu\rightarrow -r_0^2\frac{x_\mu}{ x^\mu x_\mu}$ chooses a different hyperboloid. Such a radius is used in \cite[p.~12]{Todorov2019} to modify the dimensions of $P_0$ and $K_0$ so that they can be added together. Similarly here: Recalling that $R K_\mu R =P_\mu$, a dimensional analysis shows that both terms of the operator between brackets in \Eq{eq:Rdd} are unitless.

Figure~\ref{fig:enj} shows the $c_{nj}(k)$ functions, and the corresponding photon and energy spectral densities for some values of $n$ and $j$, and $k_0=\SI{1}{\per\meter}$. Choosing different values for $k_0$ shifts the spectrum of the basis functions. Even though the countable basis is a complete orthonormal system in $\M$ {\em for any choice} of $k_0$, the spectral shift will be convenient in practical applications for roughly matching the frequency band of interest with the position of the main spectral peaks of some of the basis vectors. An alternative and physically equivalently possibility is to use $c_{nj}(k)$ or $c_{njm}(\pp)$ with $k_0=\SI{1}{\per\meter}$, but adapt the physical problem onto them by a space-time dilatation. In a dilatation, the original space-time $(\rr,t)$ is transformed into a new one $(\sss,\tau)=(\alpha \rr,\alpha t)$ with a dimensionless positive real number $\alpha$. Then, the relation between the multipolar expansion functions of the original field, $f_{jm\lambda}(k)$, and those of the transformed fields $\bar{f}_{jm\lambda}(q)$ is:
\begin{equation}
	\bar{f}_{jm\lambda}(q)=\alpha f_{jm\lambda}(\alpha q),
\end{equation}
which can be readily derived\footnote{One starts by writing the multipolar expansion of $\mathbf{\bar{E}}(\sss,\tau)$ using the explicit expansion of $\mathbf{E}(\rr,t)$ obtained from plugging \Eq{eq:mpdef} into \Eq{eq:ypans}, and finishes by performing the change of variables $k=\alpha q$.} using the known effect of a dilation on the fields \cite[Eq.~3.40a]{Fuschchich1994}: $\mathbf{\bar{E}}(\sss,\tau)=\frac{1}{\alpha^2}\mathbf{E}(\rr=\sss/\alpha,t=\tau/\alpha)$. If the original spectrum is concentrated around a given $k_0$, setting the numerical value of $\alpha=k_0$ will result in a spectrum concentrated around $k_0=\SI{1}{\meter^{-1}}$. Other practical aspects of using the $\ket{njm\lambda}$ basis in computations of light-matter interactions will be considered elsewhere.

Let us continue by studying the properties of the countable basis in $\M$.

\section{Properties of the $\ket{njm\lambda}$ in $\M$\label{sec:prop}}
 The following is proven in App.~\ref{app:proofs} using well-known integral results involving Laguerre polynomials:
\begin{equation}
	\begin{split}
		\langle \bar{\lambda}\bar{m}\bar{j}\bar{n}|njm\lambda\rangle&=\delta_{n\bar{n}}\delta_{j\bar{j}}\delta_{m\bar{m}}\delta_{\lambda\bar{\lambda}}\text{, and}\\
		\langle \lambda m j n|\op{H}|njm\lambda\rangle &=n \left(\hbar c_0k_0\right),
	\end{split}
\end{equation}
where the $\delta_{ab}$ are Kronecker deltas. Namely: The $\ket{njm\lambda}$ are orthonormal, and their energy is quantized in a physically suggestive way. The completeness of the countable basis is also proven in App.~\ref{app:proofs}.

We can therefore expand any $\ket{f}\in \M$ as:
\begin{equation}
	\label{eq:fM}
	\begin{split}
		\ket{f}&=\sum_{\lambda=\pm 1}\sum_{j=|\lambda|}^\infty \sum_{m=-j}^{j}\sum_{n=j+1}^{\infty}f_{njm\lambda}\ket{njm\lambda}\text{, and also}\\
		\ket{f}&=\sum_{\lambda=\pm 1}\sum_{n=|\lambda|+1}^\infty\sum_{j=|\lambda|}^{n-1} \sum_{m=-j}^{j}f_{njm\lambda}\ket{njm\lambda}, \text{ where}\\
		&f_{njm\lambda}=\langle \lambda m j n\ket{f}.
	\end{split}
\end{equation}
In the countable basis, and using a shortened notation, the scalar product reads:
\begin{equation}
	\braket{f|g}=\sum_{\eta}f^*_{\eta}g_{\eta}, \text{where }\eta\equiv (njm\lambda).
\end{equation}
With the finite norm condition that must be met in any Hilbert space:
\begin{equation}
	\braket{f|f}=\sum_{\eta}|f_{\eta}|^2<\infty,
\end{equation}
we find that $\M$ is isomorphic to $\ell^2$, the Hilbert space of square-summable sequences.

We will now study remarkable consequences of the properties of $c_{nj}(k)$. We will assume $k_0=\SI{1}{\per\meter}$ without loss of generality.

\subsection{Zero static fields, smoothness, and incoming/outgoing wavelets}
We note that $c_{nj}(k=0)=0$ for the allowed combinations of $n$ and $j$. This automatically excludes contributions of static fields, and it is particularly significant because it is a stronger condition that requiring a finite number of photons and a finite energy:
\begin{equation}
	\label{eq:cond}
	\braket{f|f}\le \infty,\text{ and }\braket{f|\op{H}|f}\le \infty,\text {respectively}.
\end{equation}
Finite norm is always required in a Hilbert space, and finite energy must be required on physical grounds. Any of the two conditions excludes monochromatic fields from $\M$, and the two at the same time are strong enough to exclude Lorentzian lineshapes \cite{FerCor2024}. The conditions in \Eq{eq:cond}, however, do not force $f_{jm\lambda}(k=0)=0$. For example, a particular construction using resonances with complex frequency can be built that meets $\braket{f|f}<\infty$ and $\braket{f|\op{H}|f}<\infty$, but whose expansion functions in the multipolar basis have a non-zero value at $k=0$ \cite{FerCor2024}. The countable basis excludes these latter fields from $\M$ as well, further limiting the electromagnetic fields to consider. 

A related benefit from the condition $c_{nj}(k=0)=0$, is the automatic enforcement of the exclusion of the point $k=|\pp|=0$ from the integrals in the scalar product expressions in \Eq{eq:lmsp}. One can now substitute: 
\begin{equation}
	\intdkmeasure\rightarrow \intdkZmeasure\,\text{, and }\intdpinvmz\rightarrow\intdpinv.
\end{equation}

\begin{figure}[h]
  \includegraphics[width=\linewidth]{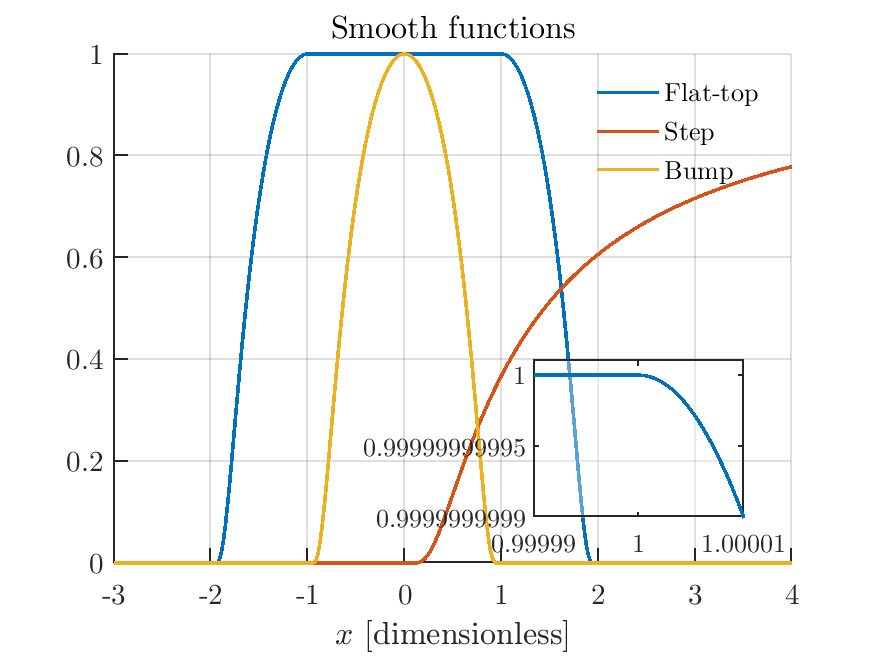}
	\caption{\label{fig:smooth}The derivatives of any order of a smooth function are continuous. The graph shows three smooth functions, equal to zero outside the following specified intervals. Step: $\exp(-1/x)$ for $x>0$; bump: $\exp\left(1/(x^2-1)\right)$ for $|x|<1$; and flat-top: $\exp\left(1-1/(1-u^2)\right)$ for $u=\text{max}(0,|x|-1)<1$. The inset shows a zoomed view of the upper-right corner of the flat-top on logarithmic axes.}
\end{figure}

\begin{figure*}
  \includegraphics[width=0.32\textwidth]{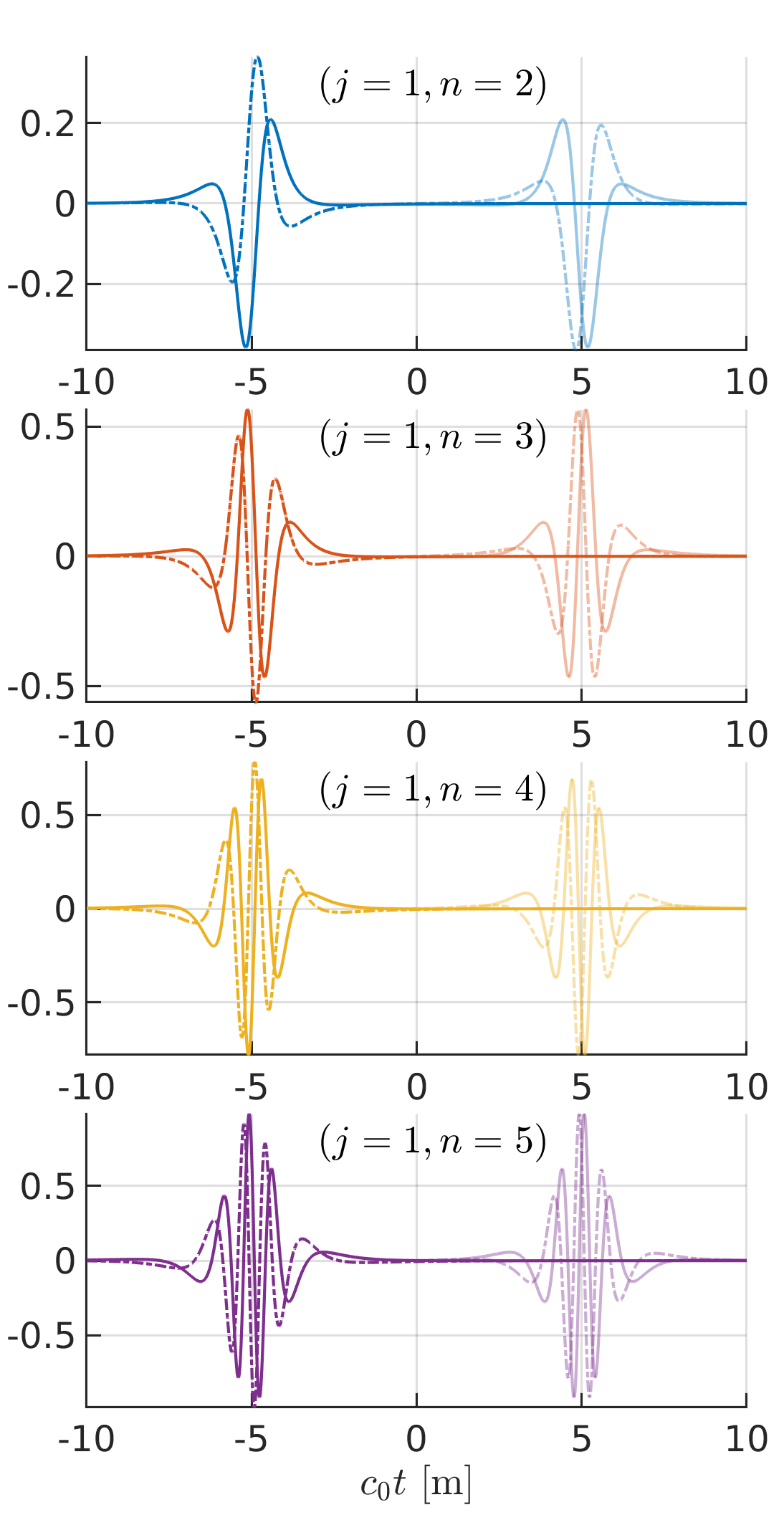}
  \includegraphics[width=0.32\textwidth]{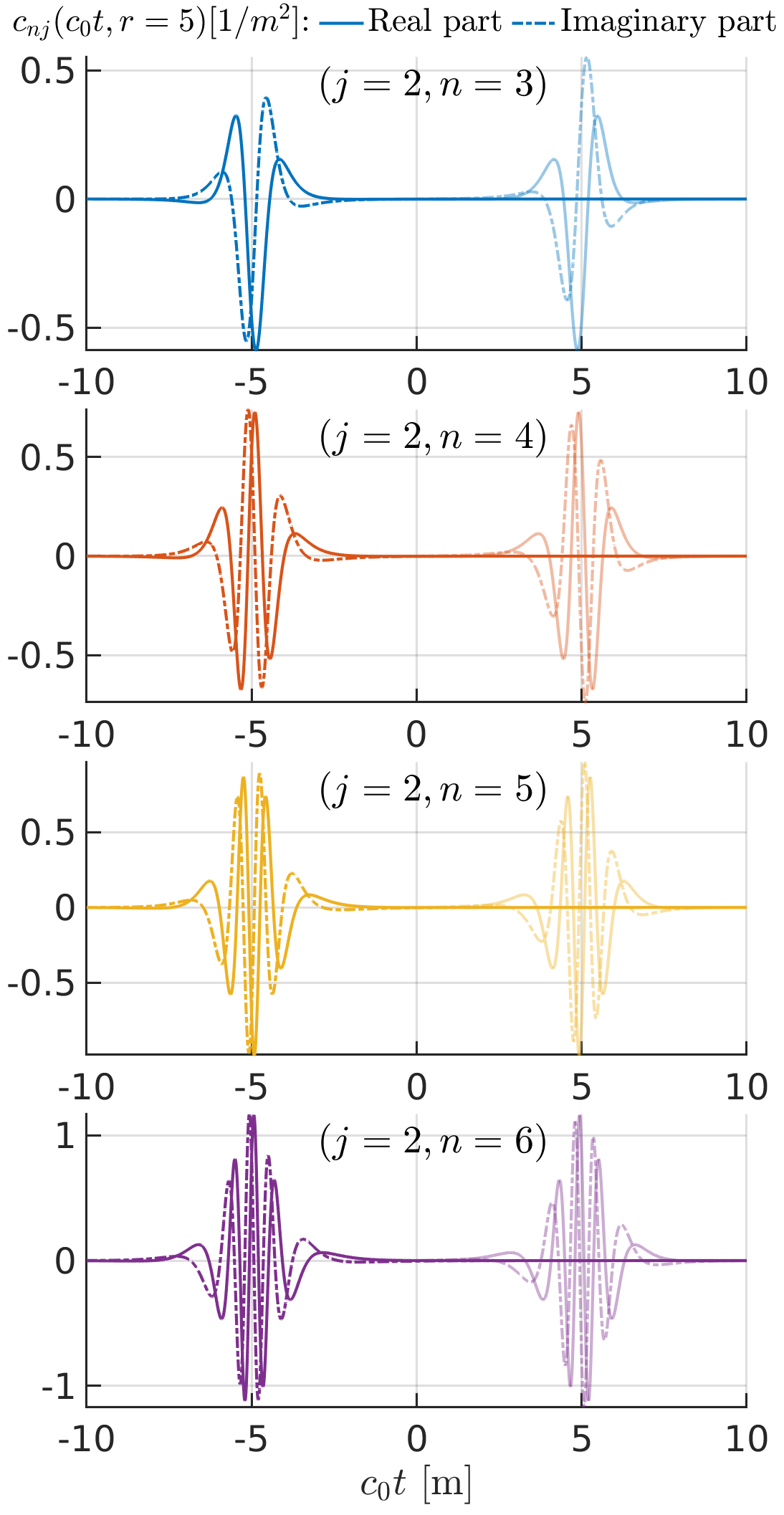}
  \includegraphics[width=0.32\textwidth]{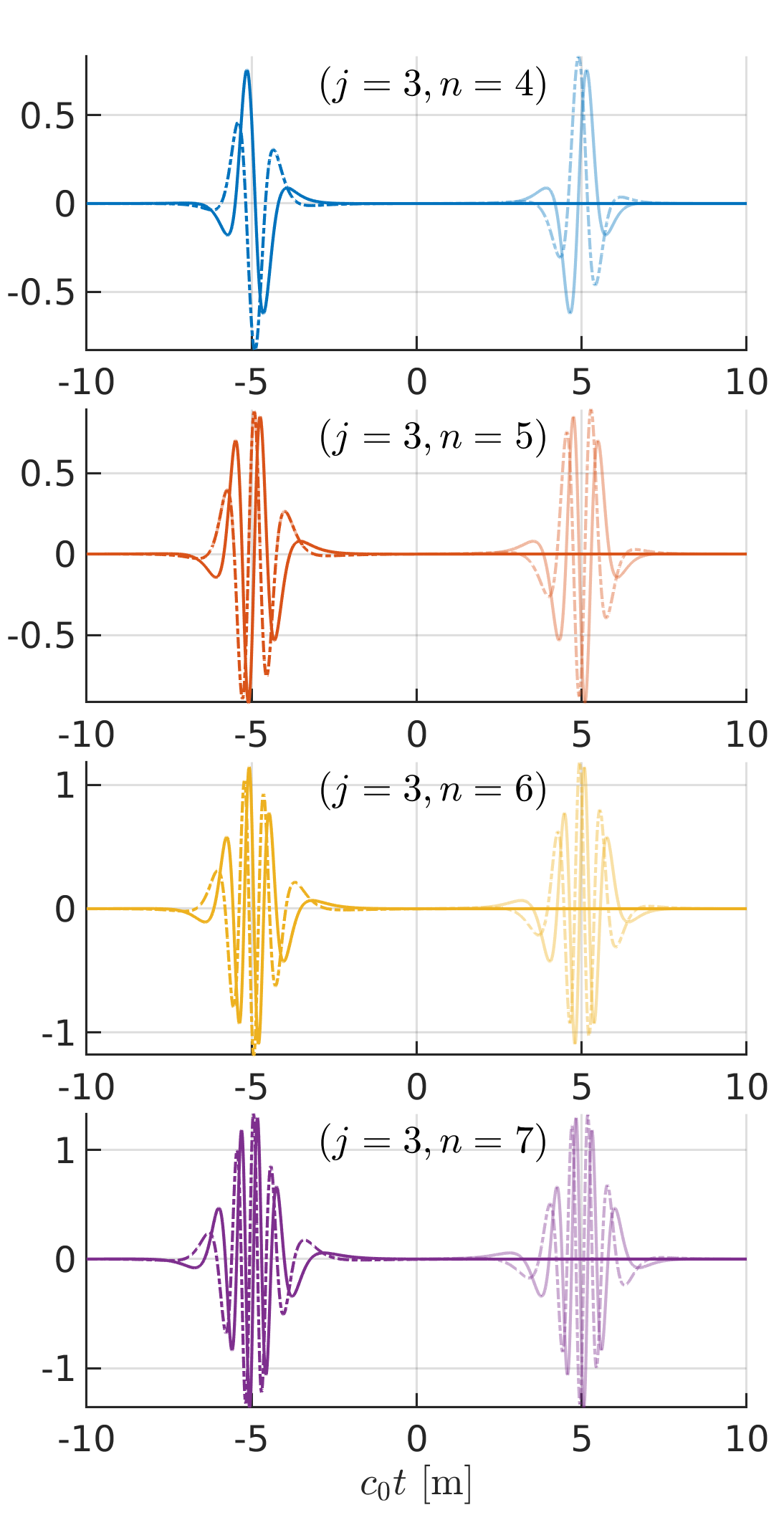}
		\caption{\label{fig:timeenj}Faint lines: Time dependence of the $l=j$ component of the electric field of $\ket{njm\lambda}$ in \Eq{eq:Ftext} at spatial radius $r=5$, denoted here by $c_{nj}(\cz t,r=5)$. Towards negative times, the faint lines are covered over by more intense lines which graph the incoming part of $c_{nj}(\cz t,r=5)$. Clearly, the incoming and outgoing parts of $c_{nj}(\cz t,r=5)$ dominate the negative and positive times, respectively. The wavelet centered around $\cz t=-5$ is almost entirely of incoming character, while the wavelet centered around $\cz t=5$ is almost entirely of outgoing character. This matches the expected behavior of an incoming spherical wave propagating from spatial infinity towards the origin and then continuing as an outgoing spherical wave propagating outwards from the origin towards spatial infinity. The value $k_0=\SI{1}{\per\meter}$ has been assumed.}
\end{figure*}

The consequences of the asymptotic behavior of the $c_{nj}(k)$ are also worth discussing. We note that, as $k\rightarrow\infty$, the $c_{nj}(k)$ decay as $(k)^u\exp(-k)$ for a fixed integer $u$. The decay is therefore faster than any polynomial decay. Appendix~\ref{app:smooth} shows that this rapid decay implies that the electromagnetic fields corresponding to the basis vectors are smooth in space and in time, that is, one can differentiate them with respect to $x,y,z,$ and $t$ an arbitrary number of times. The smooth step and smooth bump in Fig.~\ref{fig:smooth} are prototypical smooth functions of one variable. 

It is not clear whether the smoothness of the basis vectors extends to all $\ket{f}\in\M$ because a general $\ket{f}$ is an {\em infinite} linear combination of the basis vectors. In practical computations, however, a {\em finite} number of basis vectors will be considered as an approximation to a given problem, and those fields will be smooth. Smooth free Maxwell fields are produced by smooth sources \cite[Theorem~9.33]{Hintz2025}. The smoothness condition seems a very mild restriction for the sources, since the temporal and spatial scales of the ``beginnings'' and ``ends'' can be made arbitrarily small. This would be similar to scaling the dimensionless x-axis of Fig.~\ref{fig:smooth} with a chosen temporal or spatial scale. If one additionally assumes that all sources are spatially bounded and that they radiate for a finite time, then, after the radiation is finished and the field is a free field in $\M$, causality implies that the field will have compact support in space at any time instance $t$, and also that, for any point $\rr$, the field will have a compact support in time.

The time dependence of the basis fields has a remarkable double wavelet shape, where one of the wavelets is essentially of incoming character and the other of outgoing character. To show this, we start by considering the multipolar expansion of the electric field $\mathbf{E}_{njm\lambda}(\rr,t)$ corresponding to $\ket{njm\lambda}$, which can be readily obtained using \cite[Eq.~(47)]{Vavilin2023} and the $c_{nj}(k)$ coefficients in \Eq{eq:mpcoeffsnew} with $k_0=\SI{1}{\per\meter}$:
\begin{equation}
	\label{eq:Ftext}
	\begin{split}
		&\sqrt{\frac{\pi\epsz}{\cz\hbar}}\mathbf{E}_{njm\lambda}(\rr,t)=\sum_{l=j-1}^{j+1}\beta_{njl\lambda}\mathbf{Y}^l_{jm}(\rhat)\\
		&\int_{0}^\infty \mathrm{d}kk \exp(-k)(2k)^{j}L^{2j+1}_{n-j-1}(2k) k j_l(kr)\exp(-\ii \cz kt), 
	\end{split}
\end{equation}
where $\beta_{njl\lambda}$ are complex numbers, $\rhat=\rr/|\rr|$, $r=|\rr|$, and $\mathbf{Y}^l_{jm}(\rhat)$ are the vector spherical harmonics. 

The functions defined by the integral in the second line of \Eq{eq:Ftext} depend on time and on the spatial radius $r$. We denote such functions by $c_{nj}(\cz t,r)$. Figure~\ref{fig:timeenj} shows examples of such functions for $r=5$, $l=j$, and different values of $n$ and $j$. The components $l=j\pm 1$ behave similarly. We observe two wavelet-like periods centered at $\cz t \approx\mp 5$. More insight about them is gained by using the decomposition of spherical Bessel functions into spherical Hankel functions of the first and second kinds: $2j_l(kr)=h_l^1(kr)+h_l^2(kr)$. Spherical Hankel functions of the first kind produce outgoing waves, and those of the second kind produce incoming waves. The total $c_{nj}(\cz t,r=5)$, featuring the $j_l(kr)$ from \Eq{eq:Ftext}, is displayed with fainter lines in the plots. Towards negative times, the faint lines are covered over by more intense lines, which correspond to the incoming part only, obtained by the substitution $j_l(kr)\rightarrow h_l^2(kr)/2$. Then, the physical interpretation of the behavior observed in Fig.~\ref{fig:timeenj} is straightforward by considering the incoming and outgoing parts of the regular wave defined by \Eq{eq:Ftext}. The incoming wave converging towards the origin from spatial infinity reaches a given radius $r$ at around $\cz t\approx -r$, and the corresponding outgoing wave reaches the same radius $r$ again at $\cz t\approx r$.

Substituting $j_l(kr)\rightarrow h_l^1(kr)/2$ or $j_l(kr)\rightarrow h_l^2(kr)/2$ in Eq.~(\ref{eq:Ftext}) defines countable bases for outgoing or incoming fields, $\ket{njm\lambda}^{\text{out}}$ or $\ket{njm\lambda}^{\text{in}}$, which are suitable for expanding the fields involved in emission or absorption processes, respectively. The inherent polychromatic character of the basis vectors may be beneficial for describing such processes. The fields of the incoming and outgoing basis vectors are smooth in space and time, except at $\rr=\zerovec$. Such an irregular point is not much of a concern because one can always place the origin of coordinates inside the material object that motivates the consideration of incoming and outgoing waves. Then, the free fields, which are considered after their emission or absorption by the object, should vanish at all points inside the object.

\section{Conclusion and outlook\label{sec:con}}
In conclusion, a countable basis for free Maxwell fields has been identified and studied in this article. The basis vectors are defined by the integer eigenvalues of four commuting operators. Each basis vector is a polychromatic single photon wave of quantized energy. The three versions of the countable basis, regular, incoming, and outgoing, cover all the kinds of fields that are needed for the inputs and outputs of the $S$ and $T$ operators. 

These results support and motivate the use of countable bases for both the theoretical description and the practical computation of light--matter interactions. Among the expected advantages, one foresees how the double frequency integral in the current formulation of the polychromatic $S$ and $T$ operators will become a much simpler double sum over integers, avoiding the discretization of the frequency axes. Additionally, the description of polychromatic processes such as emission should benefit from the inherent polychromatic character of the basis vectors.

\begin{acknowledgments}
	I am grateful to several members of Prof.~Carsten Rockstuhl's research group for providing me with valuable feedback during a presentation of some of the subject matter contained in this article. I acknowledge funding from the Helmholtz Association via the Helmholtz program ``Materials Systems Engineering'' (MSE), and from the Deutsche Forschungsgemeinschaft (DFG, German Research Foundation) -- Project-ID 258734477 -- SFB 1173.
\end{acknowledgments}

\appendix
\section{Expansions in plane waves and multipolar fields\label{app:m}}
The electric field of a particular $\ket{f}\in\M$ is expanded into plane waves of well-defined helicity $\ket{\pp \lambda}$ as:
\begin{equation}
	\label{eq:xpans}
	\Ert=\sum_{\lambda=\pm1 }\int_{\mathbb{R}^3-\zerovec} \frac{\text{d}^3 \pp}{k} \, \ffe{\lambda} \, \ket{\pp\lambda},
\end{equation}
and the plane waves are defined as:
\begin{equation}
	\label{eq:pwdef}
	\begin{split}
		&\ket{\pp \lambda}\equiv\\
		&\sqrt{\frac{\cz\hbar}{ \epsz}}\, \frac{1}{\sqrt{2}} \frac{1}{\sqrt{(2\pi)^3}}\, k \, \mathbf{\hat{e}}_\lambda({\phat}) \exp(- \ii k\cz t ) \exp(i \pp \cdot \rr),
	\end{split}
\end{equation}
where $\cz$ is the speed of light in vacuum, $\hbar$ the reduced Planck constant, and $\epsz$ the permittivity in vacuum. We highlight the factor of $k=|\pp|$ in the definition of the plane waves, which ensures that they transform unitarily under Lorentz transformations\cite{Vavilin2023}, and the factor of $1/k$ in \Eq{eq:xpans}, which renders the volume measure $\frac{\text{d}^3 \pp}{k}$ invariant under transformations in the Poincar\'e group.

The expansion in multipoles of well-defined helicity reads:
\begin{equation}
	\label{eq:ypans}
	\begin{split}
		&{\Ert}^{\text{reg/in/out}} \equiv\\
		&\int_{>0}^\infty \text{d}k \, k \, \sum_{\lambda=\pm 1} \sum_{j=1}^{\infty} \sum_{m=-j}^j \, \FFe{\lambda} \, \ket{k j m \lambda}^{\text{reg/in/out}},
	\end{split}
\end{equation}
where $j=1,2,...$ is the multipolar degree with $j=1$ corresponding to dipoles, $j=2$ to quadrupoles, and so on, and $m=-j,-j+1,\ldots, j$ is the component of the angular momentum in the $z$ direction.

The regular, incoming, and outgoing multipoles $\ket{k j m \lambda}^{\text{reg/in/out}}$ are defined as:
\begin{widetext}
	\begin{equation}
		\begin{split}
		\label{eq:mpdef}
			&\ket{kjm\lambda}^{\text{reg}}\equiv\mathbf{S}^\text{reg}_{jm\lambda}(k,\rr,t)=\\
			&- \sqrt{\frac{\cz\hbar}{\epsilon_0}} \frac{1}{\sqrt{2\pi}} \, k \, \ii^j  \times\Big(  \exp(-\ii k \cz t)\, \mathbf{N}^{\text{reg}}_{jm}(k|\rr|, \hat{\rr}) + \lambda \,\exp(-\ii k\cz t) \,  \mathbf{M}^{\text{reg}}_{jm}(k|\rr|, \rhat ) \Big),\\
			&\ket{kjm\lambda}^{\text{in/out}}\equiv\mathbf{S}^\text{in/out}_{jm\lambda}(k,\rr,t)=\\
			&- \frac{1}{2} \sqrt{\frac{\cz\hbar}{\epsilon_0}} \frac{1}{\sqrt{2\pi}} \, k \, \ii^j\times \Big(  \exp(-\ii k \cz t)\, \mathbf{N}^{\inout}_{jm}(k|\rr|, \hat{\rr}) + \lambda \,\exp(-\ii k\cz t) \,  \mathbf{M}^{\inout}_{jm}(k|\rr|, \rhat ) \Big),
		\end{split}
	\end{equation}
\end{widetext}
where the $\mathbf{M}$ and $\mathbf{N}$ have the usual definitions (see e.g. \cite[Eqs.~(50,51)]{Vavilin2023}).

The expansions in \Eq{eq:xpans} and \Eq{eq:ypans}, together with the requirements
\begin{equation}
	\bra{\barlambda\mbar\jbar p}f\rangle=f_{\jbar\mbar\barlambda}(p),\text{ and } \bra{\barlambda\qq}f\rangle=f_{\barlambda}(\qq),
\end{equation}
impose 
\begin{equation}
	\label{eq:normswpw}
	\begin{split}
		\langle \barlambda \mbar \jbar p|k j m \lambda\rangle&=\delta_{\jbar j}\delta_{\mbar m}\delta_{\barlambda\lambda}\frac{\delta(p-k)}{k},\text{ and}\\
		\langle \barlambda \qq|\pp \lambda\rangle&=\delta_{\lambda\barlambda}\delta(\pp-\qq)k.
	\end{split}
\end{equation}

\section{Conventions in previous expressions\label{app:pexp}}
{\small
\begin{table*}[t]
\centering
\begin{tabular}{c|c|c}
	\toprule
	Ref.&	Plane wave expansion functions: {\footnotesize $k=|\pp|$, $\theta=\arccos(k_z/k)$, $\phi=\atantwo(k_y,k_x)$}  & Multipolar expansion functions\\
	\midrule
	\cite{Kastrup1970} & $\sqrt{8\frac{\nk!}{(2j+1+\nk)!}}\exp(-k)(2k)^{j}L^{2j+1}_{\nk}(2k)Y_{jm}(\phat)$& $\sqrt{8\frac{\nk!}{(2j+1+\nk)!}}\exp(-k)(2k)^{j}L^{2j+1}_{\nk}(2k)$\\
	\cite{Mack1969}& $\sqrt{\frac{(2j+1)(n-j-1)!}{(n+j)!}}\exp(-k)(2k)^{j}L^{2j+1}_{n-j-1}(2k)\exp(\ii m\phi) d^{j}_{\lambda m}(\theta)\exp(-\ii\lambda\alpha)$&not given\\
Here&$\sqrt{\frac{(2j+1)(n-j-1)!}{\pi(n+j)!}}\frac{\exp(-k/k_0)}{k_0}(2k/k_0)^{j}L^{2j+1}_{n-j-1}(2k/k_0)\exp(\ii m\phi)d^j_{m\lambda}(\theta)$& $\sqrt{\frac{4(n-j-1)!}{(n+j)!}}\frac{\exp(-k/k_0)}{k_0}(2k/k_0)^{j}L^{2j+1}_{n-j-1}(2k/k_0)$\\
\bottomrule
\end{tabular}
	\caption{\label{tab:expressions}Expressions of the expansion functions of the basis vectors of the countable basis in \cite{Kastrup1970}, \cite{Mack1969}, and in this work.}
\end{table*}
}

Table~\ref{tab:expressions} contains the expressions of the expansion functions of the basis vectors of the countable basis that appear in \cite{Kastrup1970}, \cite{Mack1969}, and in this work. The first row applies to the scalar case. It contains Eq.~(12) and Eq.~(11) from \cite{Kastrup1970} in its first and second columns, respectively. The second row, which contains Eq.~(4.4) from \cite{Mack1969}, applies to massless fields of arbitrary helicity. The third row applies to Maxwell fields. Only the expressions in the third row are dimensionally consistent thanks to the inclusion of a fixed length scale $k_0^{-1}$ (see Sec.~\ref{sec:resolution} in the main text). The factor of $1/\sqrt{\pi}$ is added on the first column of the third row to compensate different numerical factors in some expressions. The difference in the subindexes of the small Wigner matrices between the second and third rows is due to a different convention for the definition of the Wigner D-matrices: That in \cite{Gelfand1963} versus that in \cite[Chap.~7]{Tung1985}, respectively. The phase factor $\exp(-\ii\lambda\alpha)$ present in the second row has been omitted in the third. Such a factor results in the orthogonality of the different helicities through the integral in \cite[Eq.~(3.15)]{Mack1969}. Such orthogonality is already taken into account when the scalar product is written as in \Eq{eq:lmsp}, which ultimately coincides with the expression in \cite[Eq.~(3.15)]{Mack1969}. The label $n$ has a different meaning in \cite{Kastrup1970}. Referring to it as $n_{\text{Kastrup}}$, we have that: $n_{\text{Kastrup}}=n-j-1$. Then, the eigenvalues of the conformal inversion operator in \cite{Kastrup1970} should be $(-1)^{n_{\text{Kastrup}+j+1}}$, instead of $(-1)^{{n_{\text{Kastrup}}}}$. This difference also ends up equalizing the apparently different normalization factors used in \cite{Kastrup1970} and \cite{Mack1969}.

\section{Orthonormality, completeness, and energy of $\ket{njm\lambda}$\label{app:proofs}}
First, we show orthonormality:
\begin{equation}
	\begin{split}
		\langle \bar{\lambda}\bar{m}\bar{j}\bar{n}|njm\lambda\rangle&=\delta_{n\bar{n}}\delta_{j\bar{j}}\delta_{m\bar{m}}\delta_{\lambda\bar{\lambda}},\\
	\end{split}
\end{equation}
by relying on the orthonormality of the multipolar fields regarding the angular part. Using the second line of \Eq{eq:lmsp} and \Eq{eq:mpcoeffsnew}, we write:
\begin{equation}
	\begin{split}
		&\langle \bar{\lambda}\bar{m}\bar{j}\bar{n}|njm\lambda\rangle=\\
		&\sum_{\bar{\lambda}\bar{m}\bar{j}}\sum_{j m \lambda}\intdkmeasure\intdqmeasure c_{\bar{n}\bar{j}}^*(q)c_{nj}(k)\langle \bar{\lambda}\bar{m}\bar{j}q|kjm\lambda\rangle\\
		& =\delta_{\bar{j}j}\delta_{\bar{m}m}\delta_{\lambda\bar{\lambda}}\intdkmeasure c_{\bar{n}j}^*(k)c_{nj}(k).
	\end{split}
\end{equation}
where one uses $\langle \bar{\lambda}\bar{m}\bar{j}q|kjm\lambda\rangle=\frac{\delta(k-q)}{k}\delta_{\bar{j}j}\delta_{\bar{m}m}\delta_{\lambda\bar{\lambda}}$ from \Eq{eq:normswpw}. That the last line is equal to $\delta_{\bar{n}n}$ is readily deduced from the orthonormality relations of Laguerre polynomials \cite[Chap.~V,Sec.~1,Eqs.~(16$_{1,2}$)]{Sansone1959}:
\begin{equation}
	\label{eq:lag}
	\int_0^\infty \text{d}x\ \exp(-x) x^\alpha L_{s}^\alpha(x)L_{\bar{s}}^\alpha(x)=\frac{(s+\alpha)!}{s!}\delta_{s\bar{s}},
\end{equation}
and the change of variables:
\begin{equation}
	\label{eq:cov}
	\frac{x}{2}=\frac{k}{k_0},\, s=n-(j+1)\,\, \bar{s}=\bar{n}-(j+1)\,\text{, and } \alpha=2j+1.
\end{equation}
Turning now to completeness: For a given $j$, the allowed values of $n\ge j+1$ sweep all the $L^{2j+1}_{s=n-j-1}$ from $s=0,1,\ldots$. Then, the completeness of the $c_{nj}(k)$ for functions such that $\intdkmeasure |f(k)|^2<\infty$ follows from the completeness of \cite[Chap.~IV, Sec.~7]{Sansone1959}:
\begin{equation}
	\sqrt{\frac{s!}{(s+\alpha)!}}\exp(-x/2)x^{\alpha/2}L_{s}^\alpha (x),
\end{equation}
for functions such that $\int _0^\infty \text{d}x |f(x)|^2<\infty$, and the changes of variables in \Eq{eq:cov}.

Interestingly, the energy of the $\ket{njm\lambda}$ is quantized in a physically suggestive way:
\begin{equation}
	\langle \lambda m j n|\op{H}|njm\lambda\rangle =n \left(\hbar c_0k_0\right).
\end{equation}
This result can be readily reached using \Eq{eq:Hen}, \Eq{eq:mpcoeffsnew}, another well-known integral result \cite[Eq.(13)]{MathworldLaguerre}:
\begin{equation}
	\int_0^\infty \text{d}x\ \exp(-x) x^{\alpha+1} \left[L_{s}^\alpha(x)\right]^2=\frac{(s+\alpha)!}{s!}(2s+\alpha+1),
\end{equation}
and the change of variables in \Eq{eq:cov}.

\section{Smoothness of the $\ket{njm\lambda}$ fields\label{app:smooth}}
We will see here that the fields corresponding to the $\ket{njm\lambda}$ are smooth in space and time. That is, one can differentiate them with respect to $x,y,z,$ and $t$ an arbitrary number of times. We will again assume $k_0=\SI{1}{\per\meter}$ without loss of generality. 

The multipolar expansion of the electric field $\mathbf{E}_{njm\lambda}(\rr,t)$ corresponding to $\ket{njm\lambda}$ can be readily obtained using \cite[Eq.~(47)]{Vavilin2023}, and the $c_{nj}(k)$ coefficients in \Eq{eq:mpcoeffsnew} with $k_0=\SI{1}{\per\meter}$:
\begin{equation}
	\label{eq:F}
	\begin{split}
		&\sqrt{\frac{\pi\epsz}{\cz\hbar}}\mathbf{E}_{njm\lambda}(\rr,t)=\sum_{l=j-1}^{j+1}\beta_{njl\lambda}\mathbf{Y}^l_{jm}(\rhat)\\
		&\int_{0}^\infty \mathrm{d}k\underbrace{k \exp(-k)(2k)^{j}L^{2j+1}_{n-j-1}(2k) k j_l(kr)}_{F_{njl}(k)}\exp(-\ii \cz kt), 
	\end{split}
\end{equation}
where $\beta_{njl\lambda}$ are complex numbers, $\rhat=\rr/|\rr|$, $r=|\rr|$, and $\mathbf{Y}^l_{jm}(\rhat)$ are the vector spherical harmonics. 

The function $F_{njl}(k)$ has no singularities, and also meets:
\begin{equation}
	\int_{0}^\infty \mathrm{d}k |k^s F_{njl}(k)|<\infty\text{ for all } s\ge 0,
\end{equation}
because of the presence of the $\exp(-k)$ factor. This implies that the functions resulting from the integral in the second line of \Eq{eq:F} are smooth in time, since derivatives with respect to time multiply the integrand by $-\ii\cz k$.

To establish smoothness in space, we start by remarking that the $\mathbf{Y}^l_{jm}(\rhat)$ are smooth on the sphere, that is, angular derivatives of any order exist. Derivatives along the radial direction act on the integral involving $F_{njl}(k)$. The relationship:
\begin{equation}
	(2l+1)\diff{j_l(kr)}{r}=\left[lj_{l-1}(kr)-(l+1)j_{l+1}(kr)\right]k
\end{equation}
shows that upon radial differentiation of $j_l(kr)$, the two resulting terms have an extra factor of $k$. Once again, the factor $\exp(-k)$ ensures that the additional factor of $k$ does not compromise the convergence of the $\text{d}k$ integral, and the derivative is hence well-defined. By recursion, this guarantees the existence of radial derivatives of any order.

The treatment so far, based on \Eq{eq:F}, has considered the regular multipoles. The irregular versions, which are required to describe incoming and outgoing fields, are obtained \cite[Eq.~(84)]{Vavilin2023} from \Eq{eq:F} by substituting $j_l(kr)\rightarrow h_l^{r}(kr)/2$. Spherical Hankel functions of the first kind with $r=1$ produce outgoing waves, and those of the second kind with $r=2$ produce incoming waves.

The arguments given above for the regular case will still apply for showing smoothness of the outgoing and incoming versions, except at $\rr=\zerovec$. The irregularity at $\rr=\zerovec$ is not much of a concern because one can always place the origin of coordinates inside the material object that motivates the consideration of incoming and outgoing waves.


\begin{thebibliography}{45}\makeatletter
\providecommand \@ifxundefined [1]{ \@ifx{#1\undefined}
}\providecommand \@ifnum [1]{ \ifnum #1\expandafter \@firstoftwo
 \else \expandafter \@secondoftwo
 \fi
}\providecommand \@ifx [1]{ \ifx #1\expandafter \@firstoftwo
 \else \expandafter \@secondoftwo
 \fi
}\providecommand \natexlab [1]{#1}\providecommand \enquote  [1]{``#1''}\providecommand \bibnamefont  [1]{#1}\providecommand \bibfnamefont [1]{#1}\providecommand \citenamefont [1]{#1}\providecommand \href@noop [0]{\@secondoftwo}\providecommand \href [0]{\begingroup \@sanitize@url \@href}\providecommand \@href[1]{\@@startlink{#1}\@@href}\providecommand \@@href[1]{\endgroup#1\@@endlink}\providecommand \@sanitize@url [0]{\catcode `\\12\catcode `\$12\catcode
  `\&12\catcode `\#12\catcode `\^12\catcode `\_12\catcode `\%12\relax}\providecommand \@@startlink[1]{}\providecommand \@@endlink[0]{}\providecommand \url  [0]{\begingroup\@sanitize@url \@url }\providecommand \@url [1]{\endgroup\@href {#1}{\urlprefix }}\providecommand \urlprefix  [0]{URL }\providecommand \Eprint [0]{\href }\providecommand \doibase [0]{https://doi.org/}\providecommand \selectlanguage [0]{\@gobble}\providecommand \bibinfo  [0]{\@secondoftwo}\providecommand \bibfield  [0]{\@secondoftwo}\providecommand \translation [1]{[#1]}\providecommand \BibitemOpen [0]{}\providecommand \bibitemStop [0]{}\providecommand \bibitemNoStop [0]{.\EOS\space}\providecommand \EOS [0]{\spacefactor3000\relax}\providecommand \BibitemShut  [1]{\csname bibitem#1\endcsname}\let\auto@bib@innerbib\@empty
\bibitem [{\citenamefont {Lomont}\ and\ \citenamefont
  {Moses}(1964)}]{Lomont1964}  \BibitemOpen
  \bibfield  {author} {\bibinfo {author} {\bibfnamefont {J.~S.}\ \bibnamefont
  {Lomont}}\ and\ \bibinfo {author} {\bibfnamefont {H.~E.}\ \bibnamefont
  {Moses}},\ }\bibfield  {title} {\bibinfo {title} {The representations of the
  inhomogeneous lorentz group in terms of an angular momentum basis},\ }\href
  {https://doi.org/10.1063/1.1704120} {\bibfield  {journal} {\bibinfo
  {journal} {Journal of Mathematical Physics}\ }\textbf {\bibinfo {volume}
  {5}},\ \bibinfo {pages} {294} (\bibinfo {year} {1964})}\BibitemShut {NoStop}\bibitem [{\citenamefont {Kastrup}(1965)}]{Kastrup1966}  \BibitemOpen
  \bibfield  {author} {\bibinfo {author} {\bibfnamefont {H.~A.}\ \bibnamefont
  {Kastrup}},\ }\bibfield  {title} {\bibinfo {title} {Conformal group and its
  connection with an indefinite metric in hilbert space},\ }\href
  {https://doi.org/10.1103/PhysRev.140.B183} {\bibfield  {journal} {\bibinfo
  {journal} {Phys. Rev.}\ }\textbf {\bibinfo {volume} {140}},\ \bibinfo {pages}
  {B183} (\bibinfo {year} {1965})}\BibitemShut {NoStop}\bibitem [{\citenamefont {Moses}(1965{\natexlab{a}})}]{Moses1965}  \BibitemOpen
  \bibfield  {author} {\bibinfo {author} {\bibfnamefont {H.~E.}\ \bibnamefont
  {Moses}},\ }\bibfield  {title} {\bibinfo {title} {Transformation from a
  linear momentum to an angular momentum basis for particles of zero mass and
  finite spin},\ }\href {https://doi.org/10.1063/1.1704353} {\bibfield
  {journal} {\bibinfo  {journal} {Journal of Mathematical Physics}\ }\textbf
  {\bibinfo {volume} {6}},\ \bibinfo {pages} {928} (\bibinfo {year}
  {1965}{\natexlab{a}})}\BibitemShut {NoStop}\bibitem [{\citenamefont {Moses}(1965{\natexlab{b}})}]{Moses1965b}  \BibitemOpen
  \bibfield  {author} {\bibinfo {author} {\bibfnamefont {H.~E.}\ \bibnamefont
  {Moses}},\ }\bibfield  {title} {\bibinfo {title} {Transformation from a
  linear momentum to an angular momentum basis for relativistic particles of
  nonzero mass and any spin},\ }\href {https://doi.org/10.1063/1.1704766}
  {\bibfield  {journal} {\bibinfo  {journal} {Journal of Mathematical Physics}\
  }\textbf {\bibinfo {volume} {6}},\ \bibinfo {pages} {1244} (\bibinfo {year}
  {1965}{\natexlab{b}})}\BibitemShut {NoStop}\bibitem [{\citenamefont {{Moses}}(1967)}]{Moses1967}  \BibitemOpen
  \bibfield  {author} {\bibinfo {author} {\bibfnamefont {H.~E.}\ \bibnamefont
  {{Moses}}},\ }\bibfield  {title} {\bibinfo {title} {{Integration of the
  infinitesimal generators of the inhomogeneous Lorentz group and application
  to the transformation of the wave function}},\ }\href
  {https://doi.org/10.1016/0003-4916(67)90201-1} {\bibfield  {journal}
  {\bibinfo  {journal} {Annals of Physics}\ }\textbf {\bibinfo {volume} {41}},\
  \bibinfo {pages} {158} (\bibinfo {year} {1967})}\BibitemShut {NoStop}\bibitem [{\citenamefont {Mack}\ and\ \citenamefont
  {Todorov}(1969)}]{Mack1969}  \BibitemOpen
  \bibfield  {author} {\bibinfo {author} {\bibfnamefont {G.}~\bibnamefont
  {Mack}}\ and\ \bibinfo {author} {\bibfnamefont {I.}~\bibnamefont {Todorov}},\
  }\bibfield  {title} {\bibinfo {title} {Irreducibility of the ladder
  representations of u(2, 2) when restricted to the poincaré subgroup},\
  }\href {https://doi.org/10.1063/1.1664804} {\bibfield  {journal} {\bibinfo
  {journal} {J. Math. Phys.}\ }\textbf {\bibinfo {volume} {10}},\ \bibinfo
  {pages} {2078} (\bibinfo {year} {1969})}\BibitemShut {NoStop}\bibitem [{\citenamefont {Moses}(1973)}]{Moses1973b}  \BibitemOpen
  \bibfield  {author} {\bibinfo {author} {\bibfnamefont {H.~E.}\ \bibnamefont
  {Moses}},\ }\bibfield  {title} {\bibinfo {title} {Photon wave functions and
  the exact electromagnetic matrix elements for hydrogenic atoms},\ }\href
  {https://doi.org/10.1103/PhysRevA.8.1710} {\bibfield  {journal} {\bibinfo
  {journal} {Phys. Rev. A}\ }\textbf {\bibinfo {volume} {8}},\ \bibinfo {pages}
  {1710} (\bibinfo {year} {1973})}\BibitemShut {NoStop}\bibitem [{\citenamefont {Bialynicki-Birula}\ and\ \citenamefont
  {Bialynicka-Birula}(1975)}]{Birula1975}  \BibitemOpen
  \bibfield  {author} {\bibinfo {author} {\bibfnamefont {I.}~\bibnamefont
  {Bialynicki-Birula}}\ and\ \bibinfo {author} {\bibfnamefont {Z.}~\bibnamefont
  {Bialynicka-Birula}},\ }\href@noop {} {\emph {\bibinfo {title} {Quantum
  Electrodynamics}}}\ (\bibinfo  {publisher} {Pergamon, Oxford, UK},\ \bibinfo
  {year} {1975})\BibitemShut {NoStop}\bibitem [{\citenamefont {Bialynicki-Birula}(1996)}]{Birula1996}  \BibitemOpen
  \bibfield  {author} {\bibinfo {author} {\bibfnamefont {I.}~\bibnamefont
  {Bialynicki-Birula}},\ }\bibfield  {title} {\bibinfo {title} {Photon wave
  function},\ }\href@noop {} {\bibfield  {journal} {\bibinfo  {journal} {Prog.
  Optics}\ }\textbf {\bibinfo {volume} {36}},\ \bibinfo {pages} {245} (\bibinfo
  {year} {1996})}\BibitemShut {NoStop}\bibitem [{\citenamefont {Moses}(2004)}]{Moses2004}  \BibitemOpen
  \bibfield  {author} {\bibinfo {author} {\bibfnamefont {H.~E.}\ \bibnamefont
  {Moses}},\ }\bibfield  {title} {\bibinfo {title} {The role of the irreducible
  representations of the poincaré group in solving maxwell’s equations},\
  }\href {https://doi.org/10.1063/1.1704847} {\bibfield  {journal} {\bibinfo
  {journal} {Journal of Mathematical Physics}\ }\textbf {\bibinfo {volume}
  {45}},\ \bibinfo {pages} {1887} (\bibinfo {year} {2004})}\BibitemShut
  {NoStop}\bibitem [{\citenamefont {Kastrup}(2008)}]{Kastrup2008}  \BibitemOpen
  \bibfield  {author} {\bibinfo {author} {\bibfnamefont {H.}~\bibnamefont
  {Kastrup}},\ }\bibfield  {title} {\bibinfo {title} {On the advancements of
  conformal transformations and their associated symmetries in geometry and
  theoretical physics},\ }\href
  {https://doi.org/https://doi.org/10.1002/andp.200852009-1005} {\bibfield
  {journal} {\bibinfo  {journal} {Annalen der Physik}\ }\textbf {\bibinfo
  {volume} {520}},\ \bibinfo {pages} {631} (\bibinfo {year}
  {2008})}\BibitemShut {NoStop}\bibitem [{\citenamefont {Todorov}(2019)}]{Todorov2019}  \BibitemOpen
  \bibfield  {author} {\bibinfo {author} {\bibfnamefont {I.}~\bibnamefont
  {Todorov}},\ }\bibfield  {title} {\bibinfo {title} {The lure of conformal
  symmetry},\ }\href {https://doi.org/10.1007/s00220-004-1133-4} {\bibfield
  {journal} {\bibinfo  {journal} {Bulg. J. Phys.}\ }\textbf {\bibinfo {volume}
  {16}},\ \bibinfo {pages} {117} (\bibinfo {year} {2019})}\BibitemShut
  {NoStop}\bibitem [{\citenamefont {Zel'dovich}(1965)}]{Zeldovich1965}  \BibitemOpen
  \bibfield  {author} {\bibinfo {author} {\bibfnamefont {Y.~B.}\ \bibnamefont
  {Zel'dovich}},\ }\bibfield  {title} {\bibinfo {title} {The number of quanta
  as an invariant of classical electromagnetic field},\ }\href@noop {}
  {\bibfield  {journal} {\bibinfo  {journal} {Doklady Akademii Nauk SSSR (USSR)
  English translation currently published in a number of subject-oriented
  journals}\ }\textbf {\bibinfo {volume} {163}} (\bibinfo {year}
  {1965})}\BibitemShut {NoStop}\bibitem [{\citenamefont {Fernandez-Corbaton}(2025)}]{FerCor2024b}  \BibitemOpen
  \bibfield  {author} {\bibinfo {author} {\bibfnamefont {I.}~\bibnamefont
  {Fernandez-Corbaton}},\ }\bibfield  {title} {\bibinfo {title} {An algebraic
  approach to light--matter interactions},\ }\href
  {https://doi.org/https://doi.org/10.1002/apxr.202400088} {\bibfield
  {journal} {\bibinfo  {journal} {Advanced Physics Research}\ }\textbf
  {\bibinfo {volume} {4}},\ \bibinfo {pages} {2400088} (\bibinfo {year}
  {2025})}\BibitemShut {NoStop}\bibitem [{\citenamefont {Waterman}(1965)}]{Waterman1965}  \BibitemOpen
  \bibfield  {author} {\bibinfo {author} {\bibfnamefont {P.~C.}\ \bibnamefont
  {Waterman}},\ }\bibfield  {title} {\bibinfo {title} {Matrix formulation of
  electromagnetic scattering},\ }\href {https://doi.org/10.1109/PROC.1965.4058}
  {\bibfield  {journal} {\bibinfo  {journal} {Proc. IEEE}\ }\textbf {\bibinfo
  {volume} {53}},\ \bibinfo {pages} {805} (\bibinfo {year} {1965})}\BibitemShut
  {NoStop}\bibitem [{\citenamefont {Gouesbet}(2019)}]{Gouesbet2019}  \BibitemOpen
  \bibfield  {author} {\bibinfo {author} {\bibfnamefont {G.}~\bibnamefont
  {Gouesbet}},\ }\bibfield  {title} {\bibinfo {title} {T-matrix methods for
  electromagnetic structured beams: A commented reference database for the
  period 2014--2018},\ }\href
  {https://doi.org/https://doi.org/10.1016/j.jqsrt.2019.04.004} {\bibfield
  {journal} {\bibinfo  {journal} {Journal of Quantitative Spectroscopy and
  Radiative Transfer}\ }\textbf {\bibinfo {volume} {230}},\ \bibinfo {pages}
  {247} (\bibinfo {year} {2019})}\BibitemShut {NoStop}\bibitem [{\citenamefont {Mishchenko}(2020)}]{Mishchenko2020}  \BibitemOpen
  \bibfield  {author} {\bibinfo {author} {\bibfnamefont {M.~I.}\ \bibnamefont
  {Mishchenko}},\ }\bibfield  {title} {\bibinfo {title} {Comprehensive thematic
  t-matrix reference database: a 2017-2019 update},\ }\href
  {https://doi.org/https://doi.org/10.1016/j.jqsrt.2019.106692} {\bibfield
  {journal} {\bibinfo  {journal} {Journal of Quantitative Spectroscopy and
  Radiative Transfer}\ }\textbf {\bibinfo {volume} {242}},\ \bibinfo {pages}
  {106692} (\bibinfo {year} {2020})}\BibitemShut {NoStop}\bibitem [{\citenamefont {Vavilin}\ and\ \citenamefont
  {Fernandez-Corbaton}(2024)}]{Vavilin2023}  \BibitemOpen
  \bibfield  {author} {\bibinfo {author} {\bibfnamefont {M.}~\bibnamefont
  {Vavilin}}\ and\ \bibinfo {author} {\bibfnamefont {I.}~\bibnamefont
  {Fernandez-Corbaton}},\ }\bibfield  {title} {\bibinfo {title} {The
  polychromatic t-matrix},\ }\href
  {https://doi.org/https://doi.org/10.1016/j.jqsrt.2023.108853} {\bibfield
  {journal} {\bibinfo  {journal} {JQSRT}\ }\textbf {\bibinfo {volume} {314}},\
  \bibinfo {pages} {108853} (\bibinfo {year} {2024})}\BibitemShut {NoStop}\bibitem [{\citenamefont {Vavilin}\ \emph {et~al.}(2025)\citenamefont
  {Vavilin}, \citenamefont {Mazo-V\'asquez},\ and\ \citenamefont
  {Fernandez-Corbaton}}]{Vavilin2024}  \BibitemOpen
  \bibfield  {author} {\bibinfo {author} {\bibfnamefont {M.}~\bibnamefont
  {Vavilin}}, \bibinfo {author} {\bibfnamefont {J.~D.}\ \bibnamefont
  {Mazo-V\'asquez}},\ and\ \bibinfo {author} {\bibfnamefont {I.}~\bibnamefont
  {Fernandez-Corbaton}},\ }\bibfield  {title} {\bibinfo {title} {Computing the
  interaction of light pulses with objects moving at relativistic speeds},\
  }\href {https://doi.org/10.1103/PhysRevResearch.7.013132} {\bibfield
  {journal} {\bibinfo  {journal} {Phys. Rev. Res.}\ }\textbf {\bibinfo {volume}
  {7}},\ \bibinfo {pages} {013132} (\bibinfo {year} {2025})}\BibitemShut
  {NoStop}\bibitem [{\citenamefont {Whittam}\ \emph {et~al.}(2024)\citenamefont
  {Whittam}, \citenamefont {Zerulla}, \citenamefont {Krsti{\'{c}}},
  \citenamefont {Vavilin}, \citenamefont {Holzer}, \citenamefont {Nyman},
  \citenamefont {Rebholz}, \citenamefont {Fernandez-Corbaton},\ and\
  \citenamefont {Rockstuhl}}]{Whittam2024}  \BibitemOpen
  \bibfield  {author} {\bibinfo {author} {\bibfnamefont {M.~R.}\ \bibnamefont
  {Whittam}}, \bibinfo {author} {\bibfnamefont {B.}~\bibnamefont {Zerulla}},
  \bibinfo {author} {\bibfnamefont {M.}~\bibnamefont {Krsti{\'{c}}}}, \bibinfo
  {author} {\bibfnamefont {M.}~\bibnamefont {Vavilin}}, \bibinfo {author}
  {\bibfnamefont {C.}~\bibnamefont {Holzer}}, \bibinfo {author} {\bibfnamefont
  {M.}~\bibnamefont {Nyman}}, \bibinfo {author} {\bibfnamefont
  {L.}~\bibnamefont {Rebholz}}, \bibinfo {author} {\bibfnamefont
  {I.}~\bibnamefont {Fernandez-Corbaton}},\ and\ \bibinfo {author}
  {\bibfnamefont {C.}~\bibnamefont {Rockstuhl}},\ }\bibfield  {title} {\bibinfo
  {title} {Circular dichroism of relativistically--moving chiral molecules},\
  }\href {https://doi.org/10.1038/s41598-024-66443-w} {\bibfield  {journal}
  {\bibinfo  {journal} {Scientific Reports}\ }\textbf {\bibinfo {volume}
  {14}},\ \bibinfo {pages} {16812} (\bibinfo {year} {2024})}\BibitemShut
  {NoStop}\bibitem [{\citenamefont {Ambrosio}\ \emph {et~al.}(2024)\citenamefont
  {Ambrosio}, \citenamefont {{de Sarro}},\ and\ \citenamefont
  {Gouesbet}}]{Ambrosio2024}  \BibitemOpen
  \bibfield  {author} {\bibinfo {author} {\bibfnamefont {L.~A.}\ \bibnamefont
  {Ambrosio}}, \bibinfo {author} {\bibfnamefont {J.~O.}\ \bibnamefont {{de
  Sarro}}},\ and\ \bibinfo {author} {\bibfnamefont {G.}~\bibnamefont
  {Gouesbet}},\ }\bibfield  {title} {\bibinfo {title} {An approach for a
  polychromatic generalized lorenz–mie theory},\ }\href
  {https://doi.org/https://doi.org/10.1016/j.jqsrt.2023.108824} {\bibfield
  {journal} {\bibinfo  {journal} {Journal of Quantitative Spectroscopy and
  Radiative Transfer}\ }\textbf {\bibinfo {volume} {312}},\ \bibinfo {pages}
  {108824} (\bibinfo {year} {2024})}\BibitemShut {NoStop}\bibitem [{\citenamefont {von Neumann}(1955)}]{VonNeumann1955}  \BibitemOpen
  \bibfield  {author} {\bibinfo {author} {\bibfnamefont {J.}~\bibnamefont {von
  Neumann}},\ }\href@noop {} {\emph {\bibinfo {title} {Mathematical Foundations
  of Quantum Mechanics}}}\ (\bibinfo  {publisher} {Princeton University
  Press},\ \bibinfo {address} {Princeton, NJ},\ \bibinfo {year} {1955})\
  \bibinfo {note} {translated by Robert T. Beyer}\BibitemShut {NoStop}\bibitem [{\citenamefont {Streater}\ and\ \citenamefont
  {Wightman}(1989)}]{Streater1989}  \BibitemOpen
  \bibfield  {author} {\bibinfo {author} {\bibfnamefont {R.~F.}\ \bibnamefont
  {Streater}}\ and\ \bibinfo {author} {\bibfnamefont {A.~S.}\ \bibnamefont
  {Wightman}},\ }\href {http://www.jstor.org/stable/j.ctt1cx3vcq} {\emph
  {\bibinfo {title} {PCT, Spin and Statistics, and All That}}}\ (\bibinfo
  {publisher} {Princeton University Press},\ \bibinfo {year}
  {1989})\BibitemShut {NoStop}\bibitem [{\citenamefont {Earman}(2020)}]{Earman2020}  \BibitemOpen
  \bibfield  {author} {\bibinfo {author} {\bibfnamefont {J.}~\bibnamefont
  {Earman}},\ }\href {https://philsci-archive.pitt.edu/18363/} {\bibinfo
  {title} {Quantum physics in non-separable hilbert spaces}} (\bibinfo {year}
  {2020})\BibitemShut {NoStop}\bibitem [{\citenamefont {Kastrup}\ and\ \citenamefont
  {Mayer}(1970)}]{Kastrup1970}  \BibitemOpen
  \bibfield  {author} {\bibinfo {author} {\bibfnamefont {H.~A.}\ \bibnamefont
  {Kastrup}}\ and\ \bibinfo {author} {\bibfnamefont {D.~H.}\ \bibnamefont
  {Mayer}},\ }\bibfield  {title} {\bibinfo {title} {On some classes of
  solutions of the wave equation},\ }\href {https://doi.org/10.1063/1.1665194}
  {\bibfield  {journal} {\bibinfo  {journal} {Journal of Mathematical Physics}\
  }\textbf {\bibinfo {volume} {11}},\ \bibinfo {pages} {1041} (\bibinfo {year}
  {1970})}\BibitemShut {NoStop}\bibitem [{\citenamefont {Gross}(1964)}]{Gross1964}  \BibitemOpen
  \bibfield  {author} {\bibinfo {author} {\bibfnamefont {L.}~\bibnamefont
  {Gross}},\ }\bibfield  {title} {\bibinfo {title} {Norm invariance of
  {Mass‐Zero} equations under the conformal group},\ }\href
  {https://doi.org/10.1063/1.1704164} {\bibfield  {journal} {\bibinfo
  {journal} {J. Math. Phys.}\ }\textbf {\bibinfo {volume} {5}},\ \bibinfo
  {pages} {687} (\bibinfo {year} {1964})}\BibitemShut {NoStop}\bibitem [{\citenamefont {Wigner}(1939)}]{Wigner1939}  \BibitemOpen
  \bibfield  {author} {\bibinfo {author} {\bibfnamefont {E.}~\bibnamefont
  {Wigner}},\ }\bibfield  {title} {\bibinfo {title} {On unitary representations
  of the inhomogeneous lorentz group},\ }\href
  {https://doi.org/10.2307/1968551} {\bibfield  {journal} {\bibinfo  {journal}
  {Annals of Mathematics}\ }\textbf {\bibinfo {volume} {40}},\ \bibinfo {pages}
  {149} (\bibinfo {year} {1939})}\BibitemShut {NoStop}\bibitem [{\citenamefont {Bateman}(1910)}]{Bateman1910}  \BibitemOpen
  \bibfield  {author} {\bibinfo {author} {\bibfnamefont {H.}~\bibnamefont
  {Bateman}},\ }\bibfield  {title} {\bibinfo {title} {The transformation of the
  electrodynamical equations},\ }\href
  {https://doi.org/https://doi.org/10.1112/plms/s2-8.1.223} {\bibfield
  {journal} {\bibinfo  {journal} {Proceedings of the London Mathematical
  Society}\ }\textbf {\bibinfo {volume} {s2-8}},\ \bibinfo {pages} {223}
  (\bibinfo {year} {1910})}\BibitemShut {NoStop}\bibitem [{\citenamefont {Cunningham}(1910)}]{Cunningham1910}  \BibitemOpen
  \bibfield  {author} {\bibinfo {author} {\bibfnamefont {E.}~\bibnamefont
  {Cunningham}},\ }\bibfield  {title} {\bibinfo {title} {The principle of
  relativity in electrodynamics and an extension thereof},\ }\href
  {https://doi.org/https://doi.org/10.1112/plms/s2-8.1.77} {\bibfield
  {journal} {\bibinfo  {journal} {Proceedings of the London Mathematical
  Society}\ }\textbf {\bibinfo {volume} {s2-8}},\ \bibinfo {pages} {77}
  (\bibinfo {year} {1910})}\BibitemShut {NoStop}\bibitem [{\citenamefont {Kastrup}(1962)}]{Kastrup1962}  \BibitemOpen
  \bibfield  {author} {\bibinfo {author} {\bibfnamefont {H.~A.}\ \bibnamefont
  {Kastrup}},\ }\bibfield  {title} {\bibinfo {title} {Zur physikalischen
  deutung und darstellungstheoretischen analyse der konformen transformationen
  von raum und zeit},\ }\href
  {https://doi.org/https://doi.org/10.1002/andp.19624640706} {\bibfield
  {journal} {\bibinfo  {journal} {Annalen der Physik}\ }\textbf {\bibinfo
  {volume} {464}},\ \bibinfo {pages} {388} (\bibinfo {year}
  {1962})}\BibitemShut {NoStop}\bibitem [{\citenamefont {Calkin}(1965)}]{Calkin1965}  \BibitemOpen
  \bibfield  {author} {\bibinfo {author} {\bibfnamefont {M.~G.}\ \bibnamefont
  {Calkin}},\ }\bibfield  {title} {\bibinfo {title} {An invariance property of
  the free electromagnetic field},\ }\href {https://doi.org/10.1119/1.1971089}
  {\bibfield  {journal} {\bibinfo  {journal} {Am. J. Phys.}\ }\textbf {\bibinfo
  {volume} {33}},\ \bibinfo {pages} {958} (\bibinfo {year} {1965})}\BibitemShut
  {NoStop}\bibitem [{\citenamefont {Zwanziger}(1968)}]{Zwanziger1968}  \BibitemOpen
  \bibfield  {author} {\bibinfo {author} {\bibfnamefont {D.}~\bibnamefont
  {Zwanziger}},\ }\bibfield  {title} {\bibinfo {title} {Quantum field theory of
  particles with both electric and magnetic charges},\ }\href
  {https://doi.org/10.1103/PhysRev.176.1489} {\bibfield  {journal} {\bibinfo
  {journal} {Phys. Rev.}\ }\textbf {\bibinfo {volume} {176}},\ \bibinfo {pages}
  {1489} (\bibinfo {year} {1968})}\BibitemShut {NoStop}\bibitem [{\citenamefont {L{\"u}scher}\ and\ \citenamefont
  {Mack}(1975)}]{Luescher1975}  \BibitemOpen
  \bibfield  {author} {\bibinfo {author} {\bibfnamefont {M.}~\bibnamefont
  {L{\"u}scher}}\ and\ \bibinfo {author} {\bibfnamefont {G.}~\bibnamefont
  {Mack}},\ }\bibfield  {title} {\bibinfo {title} {Global conformal invariance
  in quantum field theory},\ }\href {https://doi.org/10.1007/BF01608988}
  {\bibfield  {journal} {\bibinfo  {journal} {Communications in Mathematical
  Physics}\ }\textbf {\bibinfo {volume} {41}},\ \bibinfo {pages} {203}
  (\bibinfo {year} {1975})}\BibitemShut {NoStop}\bibitem [{\citenamefont {Budinich}\ and\ \citenamefont
  {Raczka}(1993)}]{Budinich1993}  \BibitemOpen
  \bibfield  {author} {\bibinfo {author} {\bibfnamefont {P.}~\bibnamefont
  {Budinich}}\ and\ \bibinfo {author} {\bibfnamefont {R.}~\bibnamefont
  {Raczka}},\ }\bibfield  {title} {\bibinfo {title} {Global properties of
  conformally flat momentum space and their implications},\ }\href
  {https://doi.org/10.1007/BF01883768} {\bibfield  {journal} {\bibinfo
  {journal} {Foundations of Physics}\ }\textbf {\bibinfo {volume} {23}},\
  \bibinfo {pages} {599} (\bibinfo {year} {1993})}\BibitemShut {NoStop}\bibitem [{\citenamefont {Dalhuisen}(2014)}]{Dalhuisen2014}  \BibitemOpen
  \bibfield  {author} {\bibinfo {author} {\bibfnamefont {J.~W.}\ \bibnamefont
  {Dalhuisen}},\ }\emph {\bibinfo {title} {The Robinson congruence in
  electrodynamics and general relativity}},\ \href@noop {} {Ph.D. thesis},\
  \bibinfo  {school} {Leiden University} (\bibinfo {year} {2014})\BibitemShut
  {NoStop}\bibitem [{\citenamefont {Arrayás}\ \emph {et~al.}(2017)\citenamefont
  {Arrayás}, \citenamefont {Bouwmeester},\ and\ \citenamefont
  {Trueba}}]{Arrayas2017}  \BibitemOpen
  \bibfield  {author} {\bibinfo {author} {\bibfnamefont {M.}~\bibnamefont
  {Arrayás}}, \bibinfo {author} {\bibfnamefont {D.}~\bibnamefont
  {Bouwmeester}},\ and\ \bibinfo {author} {\bibfnamefont {J.}~\bibnamefont
  {Trueba}},\ }\bibfield  {title} {\bibinfo {title} {Knots in
  electromagnetism},\ }\href
  {https://doi.org/https://doi.org/10.1016/j.physrep.2016.11.001} {\bibfield
  {journal} {\bibinfo  {journal} {Physics Reports}\ }\textbf {\bibinfo {volume}
  {667}},\ \bibinfo {pages} {1} (\bibinfo {year} {2017})},\ \bibinfo {note}
  {knots in Electromagnetism}\BibitemShut {NoStop}\bibitem [{\citenamefont {Fernandez-Corbaton}(2014)}]{FerCorTHESIS}  \BibitemOpen
  \bibfield  {author} {\bibinfo {author} {\bibfnamefont {I.}~\bibnamefont
  {Fernandez-Corbaton}},\ }\emph {\bibinfo {title} {Helicity and duality
  symmetry in light matter interactions: Theory and applications}},\ \href
  {https://doi.org/10.48550/arXiv.1407.4432} {Ph.D. thesis},\ \bibinfo
  {school} {Macquarie University} (\bibinfo {year} {2014}),\ \bibinfo {note}
  {{arXiv}: 1407.4432}\BibitemShut {NoStop}\bibitem [{\citenamefont {Tung}(1985)}]{Tung1985}  \BibitemOpen
  \bibfield  {author} {\bibinfo {author} {\bibfnamefont {W.-K.}\ \bibnamefont
  {Tung}},\ }\href@noop {} {\emph {\bibinfo {title} {Group Theory in
  Physics}}}\ (\bibinfo  {publisher} {World Scientific: Singapore},\ \bibinfo
  {year} {1985})\BibitemShut {NoStop}\bibitem [{\citenamefont {Barut}\ and\ \citenamefont
  {Kleinert}(1967)}]{Barut1967}  \BibitemOpen
  \bibfield  {author} {\bibinfo {author} {\bibfnamefont {A.}~\bibnamefont
  {Barut}}\ and\ \bibinfo {author} {\bibfnamefont {H.}~\bibnamefont
  {Kleinert}},\ }\bibfield  {title} {\bibinfo {title} {Transition probabilities
  of the hydrogen atom from noncompact dynamical groups},\ }\href@noop {}
  {\bibfield  {journal} {\bibinfo  {journal} {Physical Review}\ }\textbf
  {\bibinfo {volume} {156}},\ \bibinfo {pages} {1541} (\bibinfo {year}
  {1967})}\BibitemShut {NoStop}\bibitem [{\citenamefont {Fuschchich}\ and\ \citenamefont
  {Nikitin}(1994)}]{Fuschchich1994}  \BibitemOpen
  \bibfield  {author} {\bibinfo {author} {\bibfnamefont {W.}~\bibnamefont
  {Fuschchich}}\ and\ \bibinfo {author} {\bibfnamefont {A.}~\bibnamefont
  {Nikitin}},\ }\href@noop {} {\emph {\bibinfo {title} {Symmetries of Equations
  of Quantum Mechanics}}}\ (\bibinfo  {publisher} {Allerton Press},\ \bibinfo
  {address} {New York},\ \bibinfo {year} {1994})\BibitemShut {NoStop}\bibitem [{\citenamefont {Fernandez-Corbaton}\ \emph
  {et~al.}(2024)\citenamefont {Fernandez-Corbaton}, \citenamefont {Vavilin},\
  and\ \citenamefont {Nyman}}]{FerCor2024}  \BibitemOpen
  \bibfield  {author} {\bibinfo {author} {\bibfnamefont {I.}~\bibnamefont
  {Fernandez-Corbaton}}, \bibinfo {author} {\bibfnamefont {M.}~\bibnamefont
  {Vavilin}},\ and\ \bibinfo {author} {\bibfnamefont {M.}~\bibnamefont
  {Nyman}},\ }\bibfield  {title} {\bibinfo {title} {A polychromatic theory of
  emission},\ }\href@noop {} {\bibfield  {journal} {\bibinfo  {journal} {arXiv
  preprint arXiv:2403.01319}\ } (\bibinfo {year} {2024})}\BibitemShut {NoStop}\bibitem [{\citenamefont {Hintz}(2025)}]{Hintz2025}  \BibitemOpen
  \bibfield  {author} {\bibinfo {author} {\bibfnamefont {P.}~\bibnamefont
  {Hintz}},\ }\href@noop {} {\emph {\bibinfo {title} {An introduction to
  microlocal analysis}}}\ (\bibinfo  {publisher} {Springer Cham},\ \bibinfo
  {year} {2025})\BibitemShut {NoStop}\bibitem [{\citenamefont {Gel'fand}\ \emph {et~al.}(1963)\citenamefont
  {Gel'fand}, \citenamefont {Minlos},\ and\ \citenamefont
  {Shapiro}}]{Gelfand1963}  \BibitemOpen
  \bibfield  {author} {\bibinfo {author} {\bibfnamefont {I.~M.}\ \bibnamefont
  {Gel'fand}}, \bibinfo {author} {\bibfnamefont {R.~A.}\ \bibnamefont
  {Minlos}},\ and\ \bibinfo {author} {\bibfnamefont {Z.~Y.}\ \bibnamefont
  {Shapiro}},\ }\href@noop {} {\emph {\bibinfo {title} {Representations of the
  Rotation and Lorentz Groups and their Applications}}}\ (\bibinfo  {publisher}
  {Pergamon Press},\ \bibinfo {address} {Oxford},\ \bibinfo {year}
  {1963})\BibitemShut {NoStop}\bibitem [{\citenamefont {Sansone}(1959)}]{Sansone1959}  \BibitemOpen
  \bibfield  {author} {\bibinfo {author} {\bibfnamefont {G.}~\bibnamefont
  {Sansone}},\ }\href@noop {} {\emph {\bibinfo {title} {Orthogonal
  Functions}}},\ Pure and Applied Mathematics, Vol. 9\ (\bibinfo  {publisher}
  {Interscience Publishers, Inc.},\ \bibinfo {address} {New York},\ \bibinfo
  {year} {1959})\ p.\ \bibinfo {pages} {411},\ \bibinfo {note} {revised English
  Edition}\BibitemShut {NoStop}\bibitem [{\citenamefont {{Wolfram Research}}(2024)}]{MathworldLaguerre}  \BibitemOpen
  \bibfield  {author} {\bibinfo {author} {\bibnamefont {{Wolfram Research}}},\
  }\href {https://mathworld.wolfram.com/AssociatedLaguerrePolynomial.html}
  {\bibinfo {title} {Associated laguerre polynomial}} (\bibinfo {year}
  {2024})\BibitemShut {NoStop}\end{thebibliography}
\end{document}